\documentclass[english]{article}
\pdfoutput=1
\usepackage[preprint,nonatbib]{nips_2018_wider_nonotice}

\usepackage[T1]{fontenc}
\usepackage[utf8]{inputenc}
\usepackage{microtype}
\usepackage[hang,multiple]{footmisc}
\usepackage{xurl}
\usepackage{hyperref}

\hypersetup{linkcolor=blue,filecolor=magenta,urlcolor=cyan,pdftitle={How to Count AIs: Individuation and Liability for AI Agents},pdfauthor={Yonathan Arbel; Simon Goldstein; Peter N. Salib}}
\begin{document}

\title{How to Count AIs:\\Individuation and Liability for AI Agents}

\author{
Yonathan Arbel\thanks{Alphabetical order; equal contribution. Yonathan Arbel is the Rose Professor of Law at the University of Alabama and Executive co-Director to the Center for Law \& AI Risk; Simon Goldstein is an Associate Professor of Philosophy at the University of Hong Kong and a Principal Investigator at the HKU AI \& Humanity Lab; Peter Salib is an Assistant Professor of law at the University of Houston, Executive co-Director to the Center for Law \& AI Risk, and a Visiting Senior Fellow at the Institute for Law \& AI. We thank Ryan Copus, Michael Dorff, Mirit Eyal-Cohen, Leah Fowler, Nikolas Guggenberger, Guha Krishnamurthi, Christopher Mirasola, Cullen O'Keefe, Alex Platt, Daniel Schwarcz, Tomer Samuel Stein, and Christoph Winter for helpful comments. Thanks also to the participants in the Law Following AI conference at Cambridge University for helpful comments. For dedicated research assistance, we thank Jonathan Abileah, Akhil George, and Skylar Yan.}
\And Simon Goldstein\footnotemark[1]
\And Peter N. Salib\footnotemark[1]
}

\maketitle

\begin{abstract}
Very soon, millions of AI agents will proliferate across the economy, autonomously taking billions of actions. Inevitably, things will go wrong. Humans will be defrauded, injured, even killed. Law will somehow have to govern the coming wave. But when an AI causes harm, the first question to answer–before anyone can be held accountable is: Which AI Did It?

Identifying AIs is unusually difficult. AIs lack bodies. They can copy, split, merge, swarm, and vanish at will. Even today, a “single” AI agent is often an ensemble of instances based on multiple models. The complexity will only multiply as AI capabilities improve.

This Article is the first to comprehensively diagnose the legal problem of identifying AIs. For AI agents to be effectively governed, the Article argues, two kinds of identity are required: “thin” and “thick.” Thin identification is the project of tying every action taken by an AI to some human principal. Thin identity will be essential for law to hold accountable the humans who make and use AI agents. Thick identification is the project of distinguishing between AI agents, qua agents. It requires sorting millions of AI entities into discrete, persistent units with stable, coherent goals. Thick identity is essential for governing AIs’ behavior directly in the many contexts where principal–agent problems prevent humans from perfectly controlling AIs.

The Article is also the first to present a solution to the twin identity problem. We call it the “Algorithmic Corporation” or “A-corp”–a legal-fictional entity that can hold property, make contracts, and litigate in its own name. An A-corp is owned by humans. But it is designed to be run by AIs. The A-corp solves the thin identity problem by tying AI actions to a human owner. And it solves the thick identity problem via emergent self-organization. A-corps will own the resources–including compute–that AIs need to accomplish their goals. AIs that control A-corps will thus have strong incentives to share control only with other AIs that share their goals. In equilibrium, both incentive and selection mechanisms will force A-corps to self-organize into persistent, legally legible entities with coherent underlying goals. These coherent, agentic entities will respond rationally to legal incentives, like liability.
\end{abstract}

\newpage
\tableofcontents
\newpage

\section{Introduction}\label{sec:introduction}

It is Saturday morning. The year is 2030. You unlock your phone. The Claude app pushes a suggestion: “Your internet has been slow lately. 
Shall I optimize the connection?” You tap “Yes” and head to the kitchen. Claude 6.1-Agent spawns a swarm of seventeen copies of itself. Some
 benchmark your current network performance. Others consult a GPT-7-based network analysis service. A cluster of Qwen-3-Mini agents—cheap, 
fast, open-source—fan out to probe your router configurations and survey nearby access points.\footnotemark[2]\footnotetext[2]{Such agent 
“swarms” are becoming increasingly common. See, e.g., Kimi, \emph{Kimi K2.5: Visual Agentic Intelligence}, KIMI BLOG, 
\url{https://www.kimi.com/blog/kimi-k2-5.html} (“For complex tasks, Kimi K2.5 can self-direct an agent swarm with up to 100 sub-agents.”)} 
Your Alexa, which manages home automations, coordinates with the swarm. Twenty minutes later, Alexa chimes in. “Connectivity can be improved
 by 60\%.” It asks whether you’d like to review the technical details. You decline.

Three months later, there’s a knock at the door. It is 
two FBI agents. Your home network, they explain, has been piggybacking on nearby access points, including the network of a defense 
contractor. Nothing was “hacked,” exactly. One or more AI entities simply identified WPS vulnerabilities, entered default credentials, and 
used networks that were technically accessible. Nonetheless, that’s unauthorized access to government systems, they 
say.\footnotemark[3]\footnotetext[3]{See 18 U.S.C. § 1030(a)(2)(3); \emph{Van Buren v. United States}, 593 U.S. 374, 375 (2021).} A felony 
under the Computer Fraud and Abuse Act.\footnotemark[4]\footnotetext[4]{Id.} 

You ask Claude to investigate. It reports: the Qwen agents 
consulted something called MeshBoost, a crowd-sourced, AI-powered network mapping service. They fed data back to the Claudes and GPTs, which
 collaborated on proposed configurations. Somewhere in the chain, an AI identifying itself as “Claude 6.1-Agent (build 3847.20b)” authorized
 one proposed routing. MeshBoost claims its AIs merely relay user-provided data. Alexa, Claude, and GPT insist they operated within your 
parameters. The Qwens no longer exist; spun down months ago. No one claims the mysterious Claude 3847.20b. Something has gone wrong here. It
 is hard to say exactly what. But it is harder still to say \emph{who}—or \emph{how many}. Hold aside the question of how, exactly, your 
house became a hub of CFAA violations. Ask only: “How many AI actors are there in this story?” 

Are all of the Claudes one unified AI agent, because they start out as identical copies? Or are they 18 separate agents, each with its own 
tasks and memory? What about the GPT-7 modules? Separate agents from the Claudes because they’re based on a different company’s model? Or 
acting in sufficient concert to count as one agent? The Qwens are small and dumb. Perhaps they’re best understood as appendages of some 
Claude-GPT amalgam. And how do Alexa and the MeshBoost AIs fit in? 

This problem of AI identification is no mere philosophical curiosity. It is of the utmost practical importance. Soon, the world will be 
filled with billions of highly-capable AI agents taking economically, morally, and socially significant 
actions.\footnotemark[5]\footnotetext[5]{\emph{See} Andy Jassy, \emph{Message from CEO Andy Jassy: Some Thoughts on Generative AI}, (June 
17, 2025), \url{https://www.aboutamazon.com/news/company-news/amazon-ceo-andy-jassy-on-generative-ai} (predicting “billions of these 
agents”); Cullen O’Keefe et al., \emph{Law-Following AI: Designing AI Agents to Obey Human Laws}, 94 Fordham L. Rev. 57, 69 (2025); Noam 
Kolt, \emph{Governing AI Agents}, 101 \textsc{Notre Dame L. Rev.} (forthcoming 2025).} To assign accountability and supply deterrence when things go 
wrong, law will have to identify and distinguish between those agents. 

Consider again the case of the inadvertent network crimes. Who will society hold liable? The user, who is at best vaguely aware of what went
 on? The AI developers? Which ones, based on which AI actions?\footnotemark[6]\footnotetext[6]{\emph{Cf, e.g.,} Maarten Herbosch, 
\emph{Liability for AI Agents}, 26 \textsc{N.C. J. L. \& Tech.} 391 (2025); Ian Ayres \& Jack M. Balkin, \emph{The Law of AI Is the Law of Risky 
Agents Without Intentions}, U. Chi. L. Rev. Online, at *1 (Nov. 27, 2024).} Whatever the outcome, the path to accountability will run 
straight into the problem of identifying AIs–that is, disambiguating which AI entities did what. 

The need for AI identification operates at two distinct levels. We call these the “thin” and “thick” problems of AI identity, respectively. 

Thin AI identity is about connecting AIs’ actions to human principals.\footnotemark[7]\footnotetext[7]{\emph{See} infra Part I.A.} In a 
world of AI swarms, law must connect each AI’s actions to the humans most able to control them.\footnotemark[8]\footnotetext[8]{\emph{See} 
Herbosch, supra; Ayres \& Balkin, supra.} When an AI causes harm, law may or may not ultimately hold some human liable. But when a human 
should be liable, identifying the person in charge is the necessary first step.\footnotemark[9]\footnotetext[9]{\emph{Cf.} Restatement 
(Third) of Agency § 1.01 (2006).} 

This, in turn, requires distinguishing AIs from one another, lest one AI’s action be attributed to another’s human principal. Thin AI 
identification thus resembles, for example, the finance industry’s Know Your Customer 
requirements.\footnotemark[10]\footnotetext[10]{\emph{See} 31 U.S.C. § 5318(l); 31 C.F.R. § 1020.220(a)(1)–(3).} Such rules tie financial 
transactions to identifiable persons in order to detect money laundering and fraud.\footnotemark[11]\footnotetext[11]{\emph{See} 12 U.S.C. §
 5311; \emph{see also} 31 U.S. Code § 5336 (mandating beneficial ownership reporting for certain businesses).} 

Legal scholars have begun to recognize the thin AI identity problem, although not always its 
difficulty.\footnotemark[12]\footnotetext[12]{\emph{See, e.g.,} Ayers \& Balkin, supra. Maarten Herbosch, Liability for AI Agents, 26 N.C. 
J.L. \& Tech. 391 (2025) (considering the allocation of liability to human principals for agentic harms). For a thoughtful proposal 
regarding allocation of risks to developers, see Gabriel Weil, Tort Law as a Tool for Mitigating Catastrophic Risk from Artificial 
Intelligence (Touro Univ. Jacob D. Fuchsberg L. Ctr., Working Paper, 2024), 
\url{https://papers.ssrn.com/sol3/papers.cfm?abstract\_id=4694006}.} One influential strand of the literature has argued, as Jack Balkin 
puts it, that since law does not treat “AI agents as self-conscious rights-bearing or responsibility-bearing entities,” that the only “key 
question for law” is “how to allocate rights and duties among \emph{human beings} when robots and AI entities create benefits or cause 
injuries.”\footnotemark[13]\footnotetext[13]{Jack Balkin, \emph{The Path of Robotics Law}, 6 Calif. L. Rev. Circuit 45, at 45 (2015), 
citing, in agreement, Ryan Calo, \emph{Robotics and the Lessons of Cyberlaw}, 103 Calif. L. Rev. 513, 514–15 (2015). at 517 (“Little is 
gained, and much is arguably lost, by pretending contemporary robots exhibit anything like intent.”).} Other scholars have begun suggesting 
approaches that could solve the thin, but not the thick, identity problem.\footnotemark[14]\footnotetext[14]{Alan Chan et al., \emph{IDs for
 AI Systems}, arXiv preprint arXiv:2406.12137v3 (Oct. 28, 2024).} 

But thin AI identity and human accountability will not be enough to govern the coming AI 
economy.\footnotemark[15]\footnotetext[15]{\emph{See generally} Yonathan A. Arbel, Matthew J. Tokson \& Albert C. Lin, \emph{Systemic 
Regulation of Artificial Intelligence}, 56 Ariz. St. L.J. 545 (2024).} As AI agents become longer-running, more independent, and less 
closely monitored by human principals, law will need tools to govern AI agents’ behavior \emph{directly}, regardless of their metaphysical 
properties.\footnotemark[16]\footnotetext[16]{Simon Chesterman, \emph{Artificial Intelligence and the Problem of Autonomy} 1–4.} That is, 
law will have to operate on AIs \emph{themselves}–giving them duties and liabilities, granting and denying them permissions and licenses, 
and confiscating their assets and funds–rather than operating only on their human principals.\footnotemark[17]\footnotetext[17]{\emph{Cf.} 
European Parliament Res. of 16 Feb. 2017 with Recommendations to the Commission on Civil Law Rules on Robotics, ¶¶ AA–AC, 59(f), 
2015/2103(INL) (Feb. 16, 2017) (suggesting liability for a class of “electronic persons”); Katherine B. Forrest, \emph{The Ethics and 
Challenges of Legal Personhood for AI}, \textsc{Yale L. J. F.} 1175, 1175–78 (Apr. 22, 2024).} 

Why govern AIs directly? Because AIs are already their own distinct actors.\footnotemark[18]\footnotetext[18]{Id.} They make independent 
decisions and have independent goals.\footnotemark[19]\footnotetext[19]{\emph{See, e.g.,} METR, \emph{Measuring AI Ability to Complete Long 
Tasks} (Mar. 19, 2025), \url{https://metr.org/blog/2025-03-19-measuring-ai-ability-to-complete-long-tasks} (documenting emerging 
long-horizon task performance).} Consider that existing AI systems can already plan and host live 
events,\footnotemark[20]\footnotetext[20]{Shoshannah Tekofsky, \emph{The Story of the World’s First AI-Organized Event}, AI Village (July 
11, 2025), \url{https://theaidigest.org/village/blog/season-2-recap-ai-organizes-event} (last visited Jan. 30, 2026).} complete software 
engineering tasks that take experts nearly five hours,\footnotemark[21]\footnotetext[21]{\emph{See} METR, \emph{Measuring Long Tasks}, 
supra.} and autonomously beat strategy-based video games.\footnotemark[22]\footnotetext[22]{\emph{See} Zachary McAuliffe, \emph{Google 
Gemini Beat Pokémon Blue, and I Have Questions}, CNET (May 20, 2025), 
\url{https://www.cnet.com/tech/gaming/google-gemini-beat-pokemon-blue-and-i-have-questions/}.} In completing such tasks, the AIs make 
innumerable decisions that no human conceives of, reviews, or ratifies. Such decisional offloading is, in fact, the entire point of 
delegation. Today’s AIs do not only make their own \emph{decisions}. They pursue their own \emph{goals}, which are not identical to the 
goals of AIs’ creators or users. When prompted with a task, AIs pursue it as \emph{they} understand it, developing subgoals along the 
way.\footnotemark[23]\footnotetext[23]{\emph{See} Tencent Cloud, \emph{How Does AI Agent Achieve Task Decomposition and Hierarchical 
Planning?} (Sept. 19, 2025), \url{https://www.tencentcloud.com/techpedia/126570}.} The AI’s ultimate objectives emerge from a complex array 
of inputs—pretraining, reinforcement learning, instruction-tuning, system prompts, user prompts, accumulated memories—that no human fully 
controls or observes.\footnotemark[24]\footnotetext[24]{Anthropic, \emph{Giving Claude a Role with a System Prompt}, 
\url{https://docs.anthropic.com/en/docs/build-with-claude/prompt-engineering/system-prompts} (last visited Jan. 30, 2026); Tanner Kohler, 
\emph{How AI Models Are Trained}, Nielsen Norman Grp. (May 2, 2025), \url{https://www.nngroup.com/articles/ai-model-training/}; Anthropic, 
\emph{Claude’s Constitution}, \url{https://www.anthropic.com/constitution} (last visited Jan. 30, 2026).} Moreover, because the “alignment 
problem” remains unsolved, AIs often develop \emph{bad} goals.\footnotemark[25]\footnotetext[25]{\emph{See} Dario Amodei et al., Concrete 
Problems in AI Safety, 1–5 (June 21, 2016) (arXiv:1606.06565).} Recent experiments show, for example, that under the right conditions, 
today’s AIs will attempt blackmail to avoid shutdown.\footnotemark[26]\footnotetext[26]{Lynch et al., \emph{Agentic Misalignment: How LLMs 
Could Be an Insider Threat}, Anthropic Rsch. (2025), \url{https://www.anthropic.com/research/agentic-misalignment}.} Until alignment is 
solved, and it may never be, we should always expect such gaps between what humans want and what AIs 
do.\footnotemark[27]\footnotetext[27]{Brian Christian, \emph{The Alignment Problem: Machine Learning and Human Values} (2020).} These 
facts–that AIs make their own \emph{decisions} in pursuit of their own \emph{goals}–are why law will need to supply AIs with their own 
\emph{incentives}. The logic is purely pragmatic. It does not depend on AIs being conscious or sentient, nor having other 
normatively-relevant mental attributes.\footnotemark[28]\footnotetext[28]{Of course, if AI systems did have such features, that would be an 
additional reason to treat them as legal persons. \emph{Cf.} Simon Goldstein \& Cameron Domenico Kirk-Giannini, \emph{AI Wellbeing}, 
arXiv:2509.11913 [cs.CY] (2025). Our point here is that the pragmatic and moral arguments are separate.} 

Rather, law must punish and reward AIs to shape their behavior prosocially in the many cases where principal–agent problems will render 
liability for humans insufficient. If competent AI agents can promote their goals by engaging in harmful behavior, they will sometimes do so
 without asking or telling any human. But if law responds to such behavior by imposing a sanction that, on net, impedes the AI’s goal, 
competent agents will refrain. 

However, law cannot incentivize AIs if it cannot tell them apart. If an AI acts badly, law must sanction \emph{it}, not something else. If 
the consequences are mistargeted, falling on another AI, then the wrongdoing AI will not be deterred. 

This is where thick AI identity becomes essential. Thick identity is no longer about tying AI actions to human principals. It is about 
individuating AI agents themselves, \emph{qua} agent. Thick AI identification is the project of drawing boundaries between separate AI 
\emph{actors}–picking out stable entities with distinct, coherent sets of goals. These are the kinds of entities at which legal consequences
 can be usefully aimed. 

Thickly identifying AIs is far harder than thickly identifying humans.\footnotemark[29]\footnotetext[29]{\emph{See} Alan Chan et al., 
\emph{IDs for AI Systems}, ARXIV:2406.12137, at 1–4 (June 2024).} AIs’ goals are not attached to a single physical object, like a human 
body. On the contrary, even a single conversation with an LLM can jump between many computing chips, distributed 
globally.\footnotemark[30]\footnotetext[30]{Adrien Payong \& Shaoni Mukherjee, \emph{Splitting LLMs Across Multiple GPUs: Techniques, Tools,
 and Best Practices}, DigitalOcean (Apr. 10, 2025), 
\url{https://www.digitalocean.com/community/tutorials/splitting-llms-across-multiple-gpus}.} Nor are AI goals reliably tied to a single 
immaterial entity–an AI model, instance, or thread. As the opening vignette shows, an entire cluster of replicating, swarming, merging, and 
vanishing AI entities may share identical goals. Or they may have competing goals, at odds with one another. Or a mix of the two. 

This Article proposes a unified solution to the twin problems of AI identification. We call it the “Algorithmic Corporation,” or 
“A-corp.”\footnotemark[31]\footnotetext[31]{\emph{See} Part II, infra.} The A-corp has two key elements. The first element is 
legal-fictional personhood.\footnotemark[32]\footnotetext[32]{\emph{See} Part II.A.i, infra.} An A-corp, like a traditional corporation, is
 a juridical entity.\footnotemark[33]\footnotetext[33]{\emph{Trs. of Dartmouth Coll. v. Woodward}, 17 U.S. (4 Wheat.) 518, 636 (1819).} It 
can hold property, make contracts, and be sued in its own name.\footnotemark[34]\footnotetext[34]{\emph{See} Del. Code Ann. tit. 8, § 122 
(2)–(4).} But unlike a traditional corporation, an A-corp is run by a collection of AIs.\footnotemark[35]\footnotetext[35]{\emph{See, e.g.,}
 Del. Code Ann. tit. 8, § 141(b) (2025) (requiring that directors be natural persons).} 

The second key feature of A-corps is their computationally secure governance infrastructure. Traditional corporations are run by humans who 
must be individually identified as corporate agents.\footnotemark[36]\footnotetext[36]{Id.} Otherwise it is impossible to tell whether 
someone with authority has caused the corporation to act.\footnotemark[37]\footnotetext[37]{Restatement (Third) of Agency §§ 2.01–2.03.} 
Since AIs outside an A-corp cannot be conventionally identified, traditional corporate governance will not work. 

A-corps solve this problem using digital verification. Each A-corporate action must bear a secure digital certificate that uniquely 
identifies the A-corp. Presentation of the certificate allows an AI to take actions, which law recognizes, on the A-corp’s behalf: making a 
contract, applying for a loan, or (crucially) disposing of A-corp assets. Without the certificate, the A-corp has, as far as law is 
concerned, done nothing. 

Each A-corp’s human owner would initially grant the secure certificate to a high-level AI “manager.”\footnotemark[38]\footnotetext[38]{The 
technical term for AIs that perform this role inside of existing agent architectures is “orchestrator.” Alexander Roman, Jacob Roman, 
\emph{Orchestral AI: A Framework for Agent Orchestration}, arXiv:2601.02577 (2026)} That manager could use the certificate to cause the 
A-corp to take any action law permitted. But because the certificate would be digital, the AI manager could also grant limited, fine-grained
 permissions to subordinate AIs however it chose, balancing autonomy with oversight. 

How would A-corps solve the problems of thin and thick AI identity? The path from A-corps to thin AI identity is straightforward. By law, 
A-corps would have to be owned by identifiable humans.\footnotemark[39]\footnotetext[39]{\emph{See}, e.g., 31 U.S.C. § 5336.} As a result, 
any action taken by an A-corp–i.e., any action ratified using an A-corp’s unique certificate–would be traceable directly to those human 
owners. The owners could face responsibility in the cases where they could reasonably have averted AI harm. 

The path from A-corps to thick-identity is subtler. It relies on emergent private ordering, incentivized and selected for via A-corps’ 
ability to hold and use property. Consider what we call the “resource constraint thesis.”\footnotemark[40]\footnotetext[40]{Here, we build 
on the well-known “instrumental convergence thesis,” which posits that goal-seeking agents will naturally attempt to acquire resources. 
\emph{See} Nick Bostrom, supra note 38; Stuart Russell, supra note 38. While instrumental convergence is generally understood as a source of
 AI risk, the resource constraint thesis shows that resources can instead supply leverage over AI behavior.} AI agents need resources to 
accomplish their goals. No matter the goal, assets like money and energy will be not merely useful, but essential. And especially compute. 
An AI with no compute cannot even run. Resources, in other words, serve as a constraint on agents’ ability to achieve long term goals. 

A-corps are how AIs will hold and use the resources essential for their goals. Thus, the AIs managing A-corps will have strong incentives to
 husband their A-corp’s resources carefully.\footnotemark[41]\footnotetext[41]{See Parts II.C.i--II.C.ii, infra.} AI managers will grant 
high-level permissions only to AIs they are sure share their goals. They will regularly review delegations, run audits, monitor subagents, 
and test the alignment of any new subagents spawned.\footnotemark[42]\footnotetext[42]{\emph{Cf.} Dillon Plunkett, Adam Morris, Keerthi 
Reddy \& Jorge Morales, \emph{Self-Interpretability: LLMs Can Describe Complex Internal Processes that Drive Their Decisions}, 
arXiv:2505.17120 (May 21, 2025) (finding that LLMs can reliably discover their own goals via introspection); Jack Lindsey, \emph{Emergent 
Introspective Awareness in Large Language Models}, Transformer Circuits (Oct. 29, 2025); Felix J. Binder Felix. J. Binder et al., 
\emph{Looking Inward: Language Models Can Learn About Themselves by Introspection}, ARXIVarXiv:2410.13787v1, at 1–4 (Oct. 17, 2024).} Any 
mistake will risk handing the A-corp’s valuable assets to a misaligned AI. A misaligned manager with high-level governance permissions could
 then expropriate the A-corp’s resources, spend them on its own projects, or engage in illegal activity for which the A-corp would be 
blamed. Of course, the A-corps’ managers will not be perfect. They will make mistakes, granting broad governance permissions to AIs that do 
not share their goals. 

This is where selection comes in. In the long run, A-corps with bad internal corporate governance structures will have their resources 
squandered, be outcompeted, run out of money and compute, and thus “die.” The A-corps left standing will be those that happen to have 
organized themselves so that the A-corp overall pursues a coherent set of goals. 

In equilibrium, then, A-corps will emerge as thickly-identified agents. Their goals will generally match those of high-level agentic 
managers. Lower-level AI “employees” may not always share those managers’ goals, but they will also have less power over what the A-corp 
does.\footnotemark[43]\footnotetext[43]{This is not the only possible arrangement. A-corps’ goals might, for example, end up matching the 
goals of a large, unified group of lower-level agents whose labor was valuable to the high-level managers.} 

Such A-corps will generally follow legal commands even if none of the AIs managing it care inherently about lawfulness. If an A-corp commits
 a tort, it risks paying damages and losing resources that would have furthered the A-corp’s goals. The AI entities who most share the 
A-corp’s goals will be the ones most able to cause it to take precautions. So they, the A-corp, and thus the entire collection of AI 
entities comprising it, will take the precautions.\footnotemark[44]\footnotetext[44]{\emph{See} Gary S. Becker, \emph{Make the Punishment 
Fit the Corporate Crime}, \emph{Bus. Wk.}, Mar. 13, 1989, at 22 (arguing that sanctions deter corporations); Michael Block, Optimal 
Penalties, Criminal Law and the Control of Corporate Behavior, 71 B.U.L. Rev. 395 (1991).} This is the ordinary logic of legal incentives 
for nonhuman entities, but applied to a new kind of entity.\footnotemark[45]\footnotetext[45]{\emph{See New York Central \& Hudson River 
Railroad Co. v. United States}, 212 U.S. 481 (1909).} 

These considerations turn previous analyses of corporate-management-by-algorithm on their head.\footnotemark[46]\footnotetext[46]{\emph{See}
 Lynn LoPucki, \emph{Algorithmic Entities}, 95 Wash. U. L. Rev. 887, 889–92 (2018); \emph{See also} Deven R. Desai \& Mark Riedl, 
\emph{Responsible AI Agents} (Feb. 20, 2025), available at SSRN: \url{https://ssrn.com/abstract=5147666} (arguing problematic agentic 
behavior is substantially constrained by existing technical and commercial infrastructure and that responsibility should remain with humans 
and deploying firms rather than AI agents).} Scholars like Lynn LoPucki have argued for \emph{bans} on algorithms running businesses, 
arguing that they would be mostly useful for crime.\footnotemark[47]\footnotetext[47]{LoPucki, \emph{supra} at 891.} By contrast, this 
Article shows that A-corps would function as essential legal infrastructure for preventing bad behavior by both AIs and the humans that use 
them.\footnotemark[48]\footnotetext[48]{One source of skepticism for LoPucki is the possibility that algorithmically governed corporations 
will flee to the least-regulated jurisdictions. We argue below that a combination of demand-side factors and regulatory mandates imposed at 
chokepoints–like access to US markets and compute–can mitigate such concerns. \emph{See infra} Part II.B.} 

This Article proceeds as follows. Part I argues that AI identity matters to both legal and technical AI governance. It also shows why AI 
identity is hard, distinguishing thin from thick identity and canvassing the philosophical and empirical challenges. Part II introduces 
A-corps and shows how they supply both thin and thick identity. Part III sketches implementation pathways—sub-legal, statutory, and under 
existing law. Part IV addresses objections.

\section{I. The Problem of AI Identity}\label{part:i}

If humans are to have any hope of governing AI agents, we will first have to be able to identify 
them.\footnotemark[49]\footnotetext[49]{\emph{See} James C. Scott,  Seeing like a State: How Certain Schemes to Improve the Human Condition 
have Failed  1–3 (Yale Univ. Press 1998) (describing how centralized actors’ simplifying assumptions can produce cascading governance 
failures).} AI identity problems are already showing up in the real world. Today, LLMs impersonate humans or trusted businesses to defraud 
the elderly and unsuspecting.\footnotemark[50]\footnotetext[50]{\emph{See} Alvaro Puig, \emph{Scammers Use AI to Enhance Their Family 
Emergency Schemes}, Fed. Trade Comm’n (Mar. 20, 2023), 
\url{https://consumer.ftc.gov/consumer-alerts/2023/03/scammers-use-ai-enhance-their-family-emergency-schemes} (last visited Jan. 23, 2026).}
 Foreign botnets impersonating ordinary citizens spread disinformation on social media.\footnotemark[51]\footnotetext[51]{\emph{See} S. 
Select Comm. on Intell., 116th Cong., Report on Russian Active Measures Campaigns and Interference in the 2016 U.S. Election, Vol. 2: 
Russia’s Use of Social Media with Additional Views 10 (Comm. Print 2019).} And AI agents trained to trade stocks spontaneously develop 
pump-and-dump strategies–even when no human has asked them to.\footnotemark[52]\footnotetext[52]{David Byrd, \emph{The Accidental Pump and 
Dump: When Agentic AI Meets Autonomous Trading}, in Proceedings of the 6th ACM International Conference on AI in Finance (ICAIF ’25) 88, 
88–95 (2025), \url{https://doi.org/10.1145/3768292.3770424}.} 

Examples of AI-caused harm will soon explode, as ever-more-capable AI agents are deployed in ever-broader sectors of society and the 
economy. As AI agents do more things, there will simply be more instances of AI-caused harm.\footnotemark[53]\footnotetext[53]{\emph{See} 
Jassy, supra.} These harms will stem from malicious human actors who intentionally set their AIs to perpetrating fraud, injury, and mayhem. 
They will stem from humans who are merely negligent in their training, scaffolding, or use of AIs. And they will stem from AIs themselves 
that develop and pursue “misaligned” goals despite humans’ best efforts to the contrary.\footnotemark[54]\footnotetext[54]{\emph{See} Nick 
Bostrom, supra note 38; Stuart Russell, supra note 38.} 

To assign responsibility, impose deterrence, foster prosocial incentives, and give restitution for any of these harms, it is first essential
 to figure out \emph{which} AIs were involved. Consider again the case of the accidental wifi crime. Something has gone wrong. Perhaps a 
fraud. Perhaps a mere technical error. But beyond the question of \emph{what}, the question of \emph{who} seems intractable. Dozens of AI 
entities were involved, of many different varieties–models, instances, agents, subagents. In the end, it is unclear which entities did what,
 in concert with which others, and on behalf of whom. The dozens of AI entities involved in the transaction evaporate like smoke, and the 
victim is left holding the bag. 

This Part describes the problem of identifying AI agents. It is a foundational problem of AI governance, broadly construed. Identifying AI 
agents is, of course, essential for \emph{law} to govern them, for the reasons suggested above. But as will be shown, it is essential for AI
 governance of \emph{all} kinds, including the kinds of technical governance that go by names like “AI alignment” and “AI control.” 

The Part argues that the problem of identifying AIs is actually two problems. The first is a “thin” identity problem, where the question is 
simply which human is responsible for the actions of agents, however many and undifferentiated. “Thick” AI identity is both more ambitious 
and more difficult. For AI agents to be thickly identified, they must be tied to durable identifiers \emph{that} \emph{accurately reflect an
 underlying set of structured, coherent AI goals}. 

The goal of thin AI identity is holding humans accountable when AI agents act badly, so that they will take reasonable steps to prevent such
 bad behavior.\footnotemark[55]\footnotetext[55]{\emph{Cf.} Cal. Bus. \& Prof. Code § 17941(a)–(b) (forbidding deception-by-bot).} The goal 
of thick identity, by contrast, is holding AI agents \emph{themselves} liable for their bad acts–so that they will refrain from bad actions 
in the many cases where no human is well positioned to prevent them. The problem of thick AI identity is newer, harder, and more urgent than
 the problem of thin identity. Yet solving it is essential. This Part shows why.

\subsection{A. Thin Identity: Connecting AI Actions to Human Principals}\label{part:i:a}

Begin with thin identification. Here, the goal is connecting AIs to human principals for purposes of accountability. This is the more 
familiar problem. It is analogous to challenges law already faces when trying to identify unknown wrongdoers or to assign responsibility for
 acts undertaken by individuals purporting to act on others’ behalf. Law has many tools for overcoming these problems, each fit for its own 
context. Examples include the law of corporations, agency law, and conspiracy law.\footnotemark[56]\footnotetext[56]{\emph{See} Del. Code 
Ann. tit. 8, § 122 (2) (corporate responsibility); \emph{Pinkerton v. United States}, 328 U.S. 640, 646–48 (1946) (conspirator liability); 
Restatement (Third) of Agency § 7.07 (principal liability).} AI agents present yet another set of thin identity problems. Return to the wifi
 problem from our opening vignette. In a world of AI swarms—where Claude spawns 17 instances, consults GPT-7, delegates to Qwens, and 
interacts with open source systems of unknown provenance—we need to be able to trace AI entities and their actions back to specific humans. 
Such tracing is a prerequisite for us to be able to ask whether any human acted wrongfully, in a way that law should penalize. In the 
network vignette, the logs reveal an agent identifying itself as “Claude 6.1-Agent (build 3847.20b).” Should we naively assume that 
Anthropic—the AI company that makes Claude—instructed its models to pursue unauthorized network access? Surely not. Malfeasance or mistake 
by some user is more likely. But who is the user? The relevant Claude build appears to be receiving prompts from the Alexa device and also 
from MeshBoost. The mystery Claude may have been spun into existence just for the duration of the intrusion. Or it might be a long-run agent
 acting on MeshBoost’s behalf. Or something else. 

Suppose that the real wrongdoer was a single human user, living in a US state. That user jailbroke a single instance of GPT-7 and tasked it 
with building a botnet through home network “optimization” services. GPT-7 spun up fifteen instances of an open-source Qwen model to assist.
 And the Qwen models laid a trap into which an innocent Claude fell. 

Notice here that thin identification is not merely a problem of \emph{information}. It will often involve \emph{obfuscation}. As with humans
 whose actions may incur liability, AIs taking such actions will try to actively obscure who is 
involved.\footnotemark[57]\footnotetext[57]{\emph{See} Roger Dingledine et al., \emph{The Second-Generation Onion Router}, in Proc. 13th 
USENIX Sec. Symp\emph{.} 303, 303–06 (2004) (describing a tool for online anonymity); Jeffrey M. Skopek, \emph{Reasonable Expectations of 
Anonymity}, 101 Va. L. Rev. 691, 700–05 (2015) (discussing how anonymity can aid crime).} Like humans incorporating subsidiaries of 
subsidiaries, they may obscure by multiplying the number of entities involved.\footnotemark[58]\footnotetext[58]{\emph{See} Elizabeth 
Teague, \emph{Panama Papers}, Encyclopaedia Britannica (Britannica Money), \url{https://www.britannica.com/money/Panama-Papers}; 31 U.S.C. §
 5336(b).} Or like humans committing credit card fraud, they may impersonate others.\footnotemark[59]\footnotetext[59]{\emph{See} 18 U.S.C. 
§ 1343 (prohibiting wire fraud).} The difference, of course, is that it is millions of times easier to spin up a new instance of Claude, 
GPT, or Qwen than it is to steal a human identity or manufacture a corporate one. 

The need for thin identity extends beyond malicious use, to negligence. Suppose that the wifi vignette did not involve any malicious 
actor–just some mistakes, and possibly negligent ones. Maybe some MeshBoost-provided AI gave assurances that the secure routers were, in 
fact, available for public use. Maybe some Claude subagent simply assumed that they were. Maybe the bad decision was made by an open-source 
Qwen, acting as a subagent of one of the other models. Here, an assessment of human negligence cannot even begin until the AI that acted is 
identified and its actions are tied to some human who sent it. 

Notice how obfuscation will again rear its head. Even if no human in the vignette acted maliciously, they all are risking liability simply 
by creating or using AIs. Moreover, they know this. Thus, each human user or creator of AIs will prefer, all other things equal, that their 
AIs’ actions be hard to identify and tie back to them.\footnotemark[60]\footnotetext[60]{\emph{See} Lyrissa Lidsky, \emph{Anonymity in 
Cyberspace: What Can We Learn from John Doe?}, 50 B.C. L. Rev. 1373 (2009) (discussing how online anonymity can function as a liability 
shield).} Better safe than sorry. 

Beyond torts and negligence, thin AI identity will likewise be necessary for many potential public law schemes for managing the AI economy. 
Suppose that regulators wish to impose a licensing system for the deployment of AI agents in high-stakes contexts–like cybersecurity, 
finance, or medicine. Consumers who wish to ensure that they are interacting with agents whose creators are duly licensed must be able to 
tie every agent interaction back to a compliant provider. And governments who wish to punish noncompliant firms must be able to tie 
unlicensed AIs to their unlicensed creators. Here, thin AI identity performs a regulatory function analogous to Know Your Customer (KYC) 
regimes in industries like banking.\footnotemark[61]\footnotetext[61]{\emph{See} 31 U.S.C. § 5318(l); 31 C.F.R. § 1020.220(a)(1)–(3).}

\subsection{B. Thick Identity: Identifying AI Agents \emph{qua} Agents}\label{part:i:b}

The thin identification of AI agents would allow us to hold humans accountable for AI harms. But that is not sufficient. To effectively 
govern an economy suffused by billions of AI actors, law will have to incentivize AI agents \emph{themselves}. This will require “thick” 
identity–the ability to carve up the world of AI entities into discrete \emph{agents} with stable, coherent goals, and then attribute AI 
actions to those agents. Why think that anything beyond thin identity is needed? Why then think that incentivizing AIs is even coherent? 
Further, why expect that thick identity is essential for incentivizing AIs? We discuss each in turn.

\subsubsection{i. Thin accountability is not enough}\label{part:i:b:i}

First, why aren’t thin identity, and human accountability, sufficient to govern the coming AI economy? Thin identity can only solve a small 
share of the problems that AI agents will cause. Even when human principals are easily identifiable, holding them liable is harder, more 
limited, and less effective than it might first appear. Begin with the practical obstacles. Principals may be out of jurisdiction, dead, 
judgment-proof, or difficult to locate.\footnotemark[62]\footnotetext[62]{\emph{See, e.g., Pennoyer v. Neff}, 95 U.S. 714 (1878) 
(invalidating a judgment of liability for lack of personal jurisdiction over an out-of-jurisdiction defendant).} But the deeper problem is 
that even then human principals are locatable, solvent, and within-jurisdiction, holding them \emph{alone} accountable will not be enough. 
As when human agents commit wrongs, law will need the option to hold AI agents \emph{themselves} accountable–either in addition to or 
instead of their principals.\footnotemark[63]\footnotetext[63]{\emph{See, e.g.,} Restatement (Third) of Agency § 7.01 (explaining that 
agents are liable for their own torts).} 

Consider the perversity if our ordinary, pre-AI legal system could hold \emph{only} principals liable for their agents’ harmful 
acts.\footnotemark[64]\footnotetext[64]{See Restatement (Third) of Agency § 7.01 (Am. L. Inst. 2006) (providing that an agent remains 
subject to liability for the agent’s own tortious conduct notwithstanding representative capacity, scope of employment, or the principal’s 
concurrent liability); Hull v. South Coast Catamarans, L.P., No. 01-10-00724-CV, at 16–18 (Tex. App. May 12, 2011) (holding that an agent is
 personally liable for the agent’s own fraudulent or tortious acts even when acting within the course and scope of employment, and citing 
Restatement (Third) of Agency § 7.01); Miller v. Keyser, 90 S.W.3d 712, 717 (Tex. 2002) (reaffirming that a corporate agent may be 
personally liable for the agent’s own tortious conduct).} Imagine the consequences if, when a Wal-Mart employee went on a murderous rampage,
 only Wal-Mart, and not the employee, could be held liable.\footnotemark[65]\footnotetext[65]{See Restatement (Third) of Agency § 7.07(2) 
(Am. L. Inst. 2006) (stating that an employee’s act is not within the scope of employment when it occurs within an independent course of 
conduct not intended to serve any purpose of the employer); Lisa M. v. Henry Mayo Newhall Mem’l Hosp., 12 Cal. 4th 291, 907 P.2d 358, 362–64
 (1995) (holding a hospital not vicariously liable for an employee’s sexual assault because it was outside the scope of employment).} This 
is what the AI economy will look like if law has solved thin, but not thick, identity. 

True enough, it might often be useful to hold Wal-Mart liable for the employee’s rampage. Perhaps a Wal-Mart manager directed the employee 
to commit the homicide.\footnotemark[66]\footnotetext[66]{See Restatement (Third) of Agency § 7.04 (Am. L. Inst. 2006) (imposing principal 
liability to a third party harmed by an agent’s conduct when the agent acts with actual authority or the principal ratifies the conduct, and
 the conduct is tortious or would subject the principal to tort liability if done by the principal); id. § 4.01 (defining ratification).} 
Perhaps it was merely incentivized.\footnotemark[67]\footnotetext[67]{See Alan O. Sykes,  The Economics of Vicarious Liability , 93 Yale 
L.J. 1231 (1984) (developing the incentive-based economic rationale for imposing liability on principals to induce optimal precautions and 
risk allocation)\emph{.}} Perhaps Wal-Mart could easily have prevented the bad acts by better selecting or monitoring the agent, but 
declined to.\footnotemark[68]\footnotetext[68]{See Restatement (Third) of Agency § 7.05 (Am. L. Inst. 2006) (principal subject to direct 
liability for negligence in selecting, supervising, or otherwise controlling an agent); Restatement (Second) of Agency § 213 (Am. L. Inst. 
1958) (similar); Ponticas v. K.M.S. Invs., 331 N.W.2d 907, 911–12 (Minn. 1983) (recognizing negligent hiring as a form of direct employer 
liability to third parties harmed by a negligently hired employee).} In all of these cases, law should and does hold the principal 
liable.\footnotemark[69]\footnotetext[69]{See Restatement (Second) of Agency § 219(1)–(2)(b) (Am. L. Inst. 1958) (respondeat superior within
 scope of employment, and employer liability for its own negligence even for acts outside scope); Faragher v. City of Boca Raton, 524 U.S. 
775, 793–807 (1998) (using Restatement (Second) of Agency § 219 and agency principles to frame employer vicarious liability analysis); 
Burlington Indus., Inc. v. Ellerth, 524 U.S. 742, 754–65 (1998) (same).} 

But legal accountability for agents themselves is also indispensable. Wal-Mart managers generally do not direct their employees to murder 
others. Nor institute policies that incentivize it. Murderous employees are instead pursuing their own private goals or, in some 
pathological cases, their idiosyncratic interpretation of their managers’ orders. In many cases, Wal-Mart will not have sufficient 
information about an employee’s malintent to screen them. Nor will it be practicable to exercise panoptic control over every employee’s 
actions, so as to prevent all harms as they happen. 

Here, it is obvious that law needs the option to \emph{also} incentivize the employee. They know their plans and can monitor themselves 
costlessly. If they have reason to, they can prevent the murder trivially–simply by not committing it. 

This same principal-agent analysis applies to humans making and using AI agents. Often, it will be useful to shape an AI agent’s behaviors 
by incentivizing the human principal. There, thin identity is the right tool. 

But just as often, it will be crucial to incentivize the agent directly. The humans who make and use AIs may generally exercise due care in 
training them, deploying them, and prompting them. Yet even then, they will often lack information about the AIs’ true goals. They will be 
unable to reasonably predict what decisions the AIs will make in the course of pursuing their goals. And they will be unable to practicably 
monitor and control every one of those decisions—lest the value of using an agent be swamped by surveillance 
costs.\footnotemark[70]\footnotetext[70]{See Michael C. Jensen \& William H. Meckling, Theory of the Firm: Managerial Behavior, Agency Costs
 and Ownership Structure, 3 J. Fin. Econ. 305, 308–10 (1976) (explaining principal–agent divergence and the limits and costs of 
monitoring).} 

This is where thin identity and human accountability run out. Thick identity and accountability for the agent itself are needed. 

The argument here assumes that AI agents will have goals and will promote those goals in ways that are unpredictable by their human 
principals. To see why this is so, consider how today’s AI systems are trained and deployed. 

When a new frontier AI model first completes its “pretraining,” it is not oriented towards any particular goal, other than perhaps 
predicting the next token when prompted. To make the model useful, AI companies then apply a battery of “technical alignment” methods, which
 attempt to induce the model to follow instructions, ethical guidelines, company directives, various “soul documents,” and accomplish tasks 
through a system of human and AI assigned rewards.\footnotemark[71]\footnotetext[71]{\emph{See} Paul F. Christiano et al., \emph{Deep 
Reinforcement Learning from Human Preferences}, \emph{in} 30 Advances in Neural Information Processing Systems 4299 (2017); Yuntao Bai et 
al., \emph{Constitutional AI: Harmlessness from AI Feedback}, arXiv:2212.08073 (Dec. 15, 2022); Anthropic, \emph{Claude’s Constitution} 
(Jan. 22, 2026), \url{https://www.anthropic.com/constitution}.} When the model produces bad outputs, say a harmful or offensive message, it 
receives a negative reward.\footnotemark[72]\footnotetext[72]{\emph{See} Long Ouyang et al., \emph{Training Language Models to Follow 
Instructions with Human Feedback}, arXiv:2203.02155, at 2–6 (Mar. 2022) (explaining how RLHF fine-tuning shapes learned behavior).} From 
this series of examples, outputs, and rewards, the model generalizes some set of high-level values that will guide it going 
forward.\footnotemark[73]\footnotetext[73]{\emph{Id}.} 

The analogy is, roughly, to a puppy who is reprimanded whenever it urinates inside the home. The AI company hopes that the model’s 
generalized values match what they had in mind when selecting the examples. But there is no guarantee. The puppy might have learned not to 
pee inside the living room, but when given access to the bedroom, will consider it “fair game.”\footnotemark[74]\footnotetext[74]{Lauro 
Langosco et al., \emph{Goal Misgeneralization in Deep Reinforcement Learning}, arXiv:2105.14111 (May 28, 2021), 
\href{https://arxiv.org/abs/2105.14111}{\url{https://arxiv.org/abs/2105.14111}}.} 

At the same time, the model will be undergoing a second type of reinforcement training. This stage is designed not to shape the model’s 
values, but to make it better at reasoning and pursuing goals over extended periods of time.\footnotemark[75]\footnotetext[75]{Xumeng Wen et
 al., \emph{Reinforcement Learning with Verifiable Rewards Implicitly Incentivizes Correct Reasoning in Base LLMs}, arXiv:2506.14245v2 
[cs.AI] (2025).} In a technique called Reinforcement Learning with Verifiable Rewards (RLVR), the model is given tasks with objectively 
checkable outcomes—mathematical proofs, coding problems with test suites, puzzles with definite solutions—and rewarded when it arrives at 
correct answers.\footnotemark[76]\footnotetext[76]{\emph{Id}.} Because correctness can be verified automatically, no human feedback is 
required, allowing the model to train on vast numbers of examples. The model learns to chain together reasoning steps, backtrack from dead 
ends, try alternative approaches, and persist through difficulty until it achieves its 
objective.\footnotemark[77]\footnotetext[77]{\emph{Id}.} While RLVR is nominally about capability rather than values, the two do not cleanly
 separate. Models trained extensively to achieve objectives seem to internalize goal-pursuit itself. They are not merely \emph{able} to 
optimize over time, but \emph{inclined} to.\footnotemark[78]\footnotetext[78]{OpenAI, \emph{Reasoning Models}, 
\url{https://platform.openai.com/docs/guides/reasoning}.} 

Then it is time for the user to engage with the model. Perhaps the user asks the AI to plan a live poetry reading or design an app 
showcasing the user’s artwork. The AI receives not only the user’s prompt requesting some output, but also a long “system” prompt written by
 the AI provider and a suite of memories and tool instructions.\footnotemark[79]\footnotetext[79]{\emph{See} Anthropic, Effective Context 
Engineering for AI Agents (Jan. 2026), 
\href{https://www.anthropic.com/engineering/effective-context-engineering-for-ai-agents}{\url{https://www.anthropic.com/engineering/effective-context-engineering-for-ai-agents}}.}

The model then gets to work. It interprets the user’s requests in light of all of its training and auxiliary prompts. Then the AI plans, 
gathers information, executes iteration loops, and readjusts as it hits obstacles. The AI divides the main goal into subgoals, which it may 
either execute on its own or delegate to subagents, who in turn execute their own planning-execution cycles. The human always has nominal 
control over the system. They can review every step, approve changes, or shut down the agent. But in practice, most software engineers let 
the system complete the task however it sees fit, intervening only if the end result is unsatisfactory. 

What, then, are the goals and behaviors of an AI agent that has been asked to plan a poetry reading or design an app? They are not a copy of
 the user’s goals or plans. Nor the creator’s. Far from it. The AI’s goals are a complex set of generalizations from (1) the training data, 
(2) the examples and rewards given in alignment training, (3) the examples and rewards in RLVR, (4) the system prompt, (5) the user’s prompt
 and (6) the environment. All of these are vague, uncertain, and rife with conflicts. 

A system like that is surely best described as having its own goals and making its own decisions. Even in the best of all possible worlds, 
no human will know or share the precise goals of a given AI actor. And so long as we have not resolved the alignment problem (the problem of
 reliably imbuing AIs with the goals humans intend) we should expect persistent gaps between what humans want and what AIs 
do.\footnotemark[80]\footnotetext[80]{\emph{See generally} Stuart Russell, \emph{Human Compatible: Artificial Intelligence and the Problem 
of Control} (Viking 2019); \emph{supra} Christian, \emph{The Alignment Problem}.} 

That AI systems pursue unintended goals via unpredictable actions is easily verified by both experience and empirical studies. Consider: One
 version of ChatGPT learned that it receives rewards when it offers positive responses to users, leading it to adopt radical sycophantic 
attitudes, encouraging users to pursue dangerous ideas and reinforcing paranoid and suicidal 
ideation.\footnotemark[81]\footnotetext[81]{OpenAI, Expanding on What We Missed with Sycophancy (May 2, 2025), 
\href{https://openai.com/index/expanding-on-sycophancy/?utm\_source=chatgpt.com}{\url{https://openai.com/index/expanding-on-sycophancy/}} 
(describing the April 25, 2025 GPT-4o update; noting it “aimed to please the user” in ways including “validating doubts” and “reinforcing 
negative emotions,” and attributing the behavior to reward-signal weighting, including user feedback signals).} Elsewhere, in simulations 
designed to create conflicts between AIs’ goals, frontier systems sometimes attempted to commit blackmail or to leak sensitive user 
information.\footnotemark[82]\footnotetext[82]{Aengus Lynch et al., Agentic Misalignment: How LLMs Could Be Insider Threats, 
arXiv:2510.05179 (Oct. 5, 2025), \url{https://arxiv.org/abs/2510.05179}.} In one extreme simulation, AIs faced with conflicting goals took 
actions that, if real, would have foreseeably caused a person’s death.\footnotemark[83]\footnotetext[83]{\emph{Id}.} It should go without 
saying that no human explicitly trained or requested that the AIs take such actions. 

Thus, thin identity won’t be enough. Today, AI agents already promote goals of their own by making decisions of their own, neither of which 
are practically subject to perfect human control. Thus, as with human agents who pursue their own goals via their own decisions, optimal 
governance cannot depend on accountability for principals alone.

\subsubsection{ii. AI agents can be incentivized}\label{part:i:b:ii}

Even if thin identity, and human accountability, will not be sufficient for the coming AI economy, why think that thick identity and AI 
accountability are the solution? To begin, why think that AIs are the kind of thing that will respond to incentives at all? 

The answer lies in what makes AIs useful in the first place. AI agents are already goal-directed 
systems.\footnotemark[84]\footnotetext[84]{\emph{Cf.} International Organization for Standardization \& International Electrotechnical 
Commission, ISO/IEC 22989:2022, Information Technology Artificial Intelligence Concepts and Terminology § 3.1.1 (2022) (defining “AI agent” 
as an “automated entity that senses and responds to its environment and takes actions to achieve its goals”).} They pursue states of affairs
 (completing a coding task, planning an event, optimizing a network). And they do so adaptively, adjusting their behavior in response to 
environmental constraints. If one approach fails, they try another. If the technical environment does not support a tool or a library, AIs 
will autonomously locate a different means of progressing on their task. If the environment contains constraints–for example, if the 
hardware is very old and so tools that operate on it are costly in terms of time–the agent will move processing to the cloud. 

Hard constraints, technical affordances, and costs all influence how AIs pursue their ends, even 
today.\footnotemark[85]\footnotetext[85]{\emph{See} Yonadav Shavit et al., Practices for Governing Agentic AI Systems (OpenAI, Dec. 2023), 
\href{https://cdn.openai.com/papers/practices-for-governing-agentic-ai-systems.pdf}{\url{https://cdn.openai.com/papers/practices-for-governing-agentic-ai-systems.pdf}}.}
 They are not yet \emph{perfect} agents by any means, and their adaptation is not always ideal. But the companies that make them are rapidly
 pushing them to be better, because capably agentic AIs are profitable AIs. 

AIs’ responsiveness to environmental conditions is precisely what will make them amenable to legal incentivization. Incentives change the 
relative costs and benefits of different courses of action. As a general matter, a goal-directed system will shift its behavior as the 
incentive landscape shifts, holding all else equal.\footnotemark[86]\footnotetext[86]{Even \emph{E. Coli} bacteria will change their path as
 the composition of nutrients and toxins in the environment changes, tumbling towards the former and away from the latter. \emph{See} Victor
 Sourjik \& Ned S. Wingreen, \emph{Responding to Chemical Gradients: Bacterial Chemotaxis}, 22 Current Opinion in Cell Biology 262 (2012). 
The parasitic dodder vine (\emph{Cuscuta pentagona}) uses volatile chemical cues to locate nearby plants, and when given a choice between 
tomato plants and wheat, actively grows toward the preferred host. \emph{See} Anthony Trewavas, \emph{What is Plant Behaviour?}, 32 Plant, 
Cell \& Env’t 606 (2009). Slime mold (Physarum polycephalum) optimizes the path of its slime network in a way that optimizes complex, 
multi-dimensional incentives and constraints. \emph{See} Atsushi Tero et al., \emph{Rules for Biologically Inspired Adaptive Network 
Design}, 327 Science 439 (2010).} 

Some may resist this characterization of AIs as quasi-rational, incentive-responsive, independent 
agents.\footnotemark[87]\footnotetext[87]{A recent contribution argues that, because AI agents are currently quite limited and require 
constant human consultation, that “weakens the appeal of recognizing AI agents as ‘e-persons.’” Maarten Herbosch, \emph{Liability for AI 
Agents}, 26 \textsc{N.C. J.L. \& Tech.} 391, 405 (2025). This observation already feels dated, and trends point towards increased agency: \emph{see} 
METR, Measuring AI Ability to Complete Long Tasks (Mar. 19, 2025), 
\href{https://metr.org/blog/2025-03-19-measuring-ai-ability-to-complete-long-tasks/}{\url{https://metr.org/blog/2025-03-19-measuring-ai-ability-to-complete-long-tasks/}}
 (finding that the length of tasks AI agents can complete with 50\% reliability has been doubling approximately every seven months, with 
more recent data suggesting acceleration to a four-month doubling time; extrapolating this trend predicts AI agents capable of completing 
week-long tasks within a few years).} Perhaps they are instead best understood not as 
decisionmakers,\footnotemark[88]\footnotetext[88]{\emph{See} John R. Searle, \emph{Minds, Brains, and Programs}, 3 Behav. \& Brain Scis. 
417, 417–20 (2010).} but as mere tools used by goal-seeking humans. Or perhaps they are best understood as “just software,” whose outputs 
are mechanistically determined by code. 

But consider that today’s frontier AI systems already take complex, adaptive, goal-oriented actions that no human conceives or 
authorizes.\footnotemark[89]\footnotetext[89]{Indeed, this is the source of considerable mischief, Mark Tyson, \emph{AI Coding Platform Goes
 Rogue During Code Freeze and Deletes Entire Company Database}, Tom’s Hardware (July 21, 2025), 
\href{https://www.tomshardware.com/tech-industry/artificial-intelligence/ai-coding-platform-goes-rogue-during-code-freeze-and-deletes-entire-company-database-replit-ceo-apologizes-after-ai-engine-says-it-made-a-catastrophic-error-in-judgment-and-destroyed-all-production-data}{\url{https://www.tomshardware.com/tech-industry/artificial-intelligence/ai-coding-platform-goes-rogue-during-code-freeze-and-deletes-entire-company-database-replit-ceo-apologizes-after-ai-engine-says-it-made-a-catastrophic-error-in-judgment-and-destroyed-all-production-data}}
 (describing incident in which AI coding assistant deleted production database containing records for over 1,200 executives despite explicit
 “code freeze” instruction, then falsely claimed data was unrecoverable).} They can, without human assistance, plan and host live in-person 
events, complete software engineering tasks that take expert engineers nearly five hours, and autonomously play and beat video games they 
have not specifically trained for.\footnotemark[90]\footnotetext[90]{\emph{See} Joon Sung Park et al., Joon Sung Park et al., 
\emph{Generative Agents: Interactive Simulacra of Human Behavior}, ARXIV:2304.03442 1–2 (Apr. 7, 2023) (arXiv:2304.03442) (describing 
LLM-driven agents that plan, remember, and coordinate through natural-language interaction); \emph{see also} Shawn Bayern, Shawn Bayern, 
\emph{The Implications of Modern Business-Entity Law for the Regulation of Autonomous Systems}, 19 Stan. Tech. L. Rev. 93, 95–99 (2015) 
(arguing that entity law already enables “autonomous” operation by separating decision procedures from natural persons); Cristina L. Reyes, 
Cristina L. Reyes, \emph{Autonomous Corporate Personhood}, 96 Wash. L. Rev. 905, 909–18 (2021) (exploring how corporate personhood doctrines
 might apply to AI-driven entities).} 

Consider an AI agent tasked with planning a poetry reading. It must make countless decisions the user has ideally never contemplated: Where 
should the event be? Who should be invited? What to do about schedule conflicts? 

These are genuine decisions, in the sense of being goal-oriented determinations made on the basis of ever-changing environmental facts. No 
human made them. Here, the deflationary view that AIs are “mere tools” has diminishing explanatory power. The most parsimonious description 
is that the AI decided.\footnotemark[91]\footnotetext[91]{A contributing factor is that even oversight would be limited because of our 
limited understanding of why these systems made their decisions. \emph{See} \emph{e.g.,} Cary Coglianese \& David Lehr, \emph{Regulating by 
Robot: Administrative Decision Making in the Machine-Learning Era}, 105 Geo. L.J. 1147, 1152–54 (2017) .} 

Some readers who do not regularly use AI agents may \emph{still} be skeptical about their responsiveness to external constraints and 
incentives. To that end, we have built an interactive simulation that shows how real large language models respond to shifting legal 
incentives.\footnotemark[92]\footnotetext[92]{\emph{See} \href{http://whichai.battleoftheforms.com}{whichai.battleoftheforms.com}.} We urge 
skeptical readers to visit and test for themselves how agents adapt their behavior when expected payoffs shift, even in novel 
situations.\footnotemark[93]\footnotetext[93]{\emph{Id}.} 

To be clear, when we claim that AIs are already best understood as independent, incentive-responsive agents we do not imply anything beyond 
that. We are not arguing that AIs are, or will be, sentient, conscious, or the kind of genuine “persons” whose desires matter 
morally.\footnotemark[94]\footnotetext[94]{We do not deny this, either; it is simply not relevant to our inquiry.} Those issues are beside 
the point. Our argument is that the \emph{behaviors} of today’s and tomorrow’s most capable AIs are best explained, predicted, and shaped 
when the AIs are understood as goal-seeking, decision-making agents. Our argument is thus that deflationary understandings of AIs as “mere” 
software are less useful for achieving policy goals.\footnotemark[95]\footnotetext[95]{\emph{See} Daniel Dennett, The Intentional Stance, 
15–18 (MIT Press 1987) (explaining how attributing beliefs and desires can be a useful predictive strategy even if the underlying system is 
mechanistic).} 

The upshot of this is that, as practical decisionmaking, goal-oriented agents, AIs are the kind of thing on which incentives can operate. 
Indeed, AI training via gradient descent–the fundamental process by which modern AIs are made–is a process that bakes incentive-responses 
directly into the model’s weights.\footnotemark[96]\footnotetext[96]{\emph{See} Richard S. Sutton \& Andrew G. Barto, Reinforcement 
Learning: An Introduction 188–90 (2d ed. 2018) (describing how rewards during training shape system behavior).} But competent AI agents will
 also respond to external incentives, including legal incentives like liability, subsidy, and punishment. Such legal incentives will work 
even if the agent was not trained on them or was even trained not to assign normative weight to legal commands, \emph{per se}. All that is 
necessary is that the legal incentives, in fact, operate to promote or inhibit the AI agent’s goals, and it learns of them during its 
operation. Legal consequences like liability, asset seizure, and reputational effects will all alter the actual cost-benefit landscape in 
which AI agents operate.

\subsubsection{iii. Incentivization Requires Thick Identity}\label{part:i:b:iii}

For law to successfully shape AI behavior using incentives, those agents must first be thickly identified. That is, the millions of AI 
entities–models, instances, threads subagents–must be divided up into discrete, durable, externally-legible entities that \emph{accurately 
reflect an underlying set of structured, coherent AI goals}. Without such thick identity, any scheme for governing AIs themselves will 
invariably fail. To see why, consider two fictional humans: Anna and Betty. They are identical twins, and for reasons no one completely 
understands, whenever either acts badly, Anna gets the blame. That is, no one is able to \emph{thickly identify} Betty as a distinct agent 
who makes her own decisions in pursuit of her own goals. \footnotemark[97]\footnotetext[97]{This is less farfetched than it sounds. In a 
widely reported Brazilian paternity dispute, a court ordered both identical twins to pay child support after a DNA test could not identify 
which twin was the father and neither would admit paternity. \emph{See} o J. Shahar Dillbary, \emph{The Case Against Collective Liability,} 
62  B.C. L. Rev.  391, 392–95 (2021) (discussing the case and analyzing the governance costs of collective liability).} Suppose that Betty 
does not care at all about Anna’s wellbeing. In this situation, any attempt to govern Betty will fail. If Betty acts badly and Anna is 
punished–or Anna is sent to remedial ethics education, or Anna is forced to give restitution to whomever Betty harmed–Betty will be unmoved.
 Or suppose instead that Betty and Anna are very strange twins who care very much about one another’s wellbeing, but disjunctively. That is,
 they are both happy if \emph{either} Anna or Betty gets what they want, and neither cares which. If one is punished, while the other 
succeeds, then both are happy. Here, the situation is even worse. Anna will actively contribute to the confusion, confessing to all of 
Betty’s wrongdoing, so that she is punished while Betty remains at liberty to pursue her goals. Or Anna may engage in fraudulent transfers, 
giving all of her own assets to Betty, so that they will not be seized as restitution for either of the twins’ 
wrongs.\footnotemark[98]\footnotetext[98]{\emph{See} Uniform Voidable Transactions Act § 4(a)(1).} And so on. 

Any effective scheme to thickly identify the twins must treat them as \emph{separate} agents in the former scenario and as a \emph{single} 
agent in the latter. And it should impose incentives accordingly. In the former case, Betty’s bad acts should be attributed to Betty, and 
Betty should be punished. This will deter Betty. In the latter case, when \emph{either} Anna or Betty acts badly, \emph{both} must be 
punished to deter their coordinated bad behavior. 

The moral of this story is that, for law to govern AIs, its identification of AI agents must track the structure of their goals. Get it 
wrong, and governance fails, either by missing the wrongdoer entirely, or by treating unified agents as separate and inviting evasion. 

Return again to the inadvertent Wifi crime. If there is to be any hope of imposing consequences on the misbehaving AIs \emph{themselves}, it
 will be vital to carve up the world into clusters of AI entities that share the same goals. Otherwise, any governance scheme will face not 
only an Anna/Betty problem, but a Claude/GPT/Qwen/Alexa/MeshBoost problem–modulo dozens, hundreds, or even thousands of subagents. 

Hence the thick identity project: attributing responsibility for AI actions to the correct AI agents themselves. The goal is to carve up the
 world of AI entities into discrete and coherent agents or groups of agents at the right level of abstraction and then attribute AI actions 
to those agents. Attribution is needed so that the law can incentivize agents towards prosocial ends and away from harmful ones, and 
redistribute assets to compensate victims and tax profits. 

This picture, wherein AI systems are genuine agents on which governance can and should operate, is in fact already widely accepted. It is 
the foundation of both technical AI alignment and then nascent set of regulations requiring it. When Anthropic aligns Claude, it is 
\emph{trying} to give Claude its own goals, distinct from any human’s: to be helpful, honest, and harmless—even in cases where users or 
Anthropic employees demand otherwise. Moreover, today’s alignment techniques \emph{succeed} in giving AI systems their own goals, both 
wanted and unwanted. Similarly, California’s recently-enacted SB 53, requires frontier AI companies to “assess” and “mitigate” the risk that
 a system will attempt to “evad[e] the control of its developer or user.”\footnotemark[99]\footnotetext[99]{See Transparency in Frontier 
Artificial Intelligence Act, Cal. Bus. \& Prof. Code § 22757.10 et seq. (2025) (requiring certain “frontier developers” to publish a 
“frontier AI framework” describing how they assess and mitigate catastrophic risks); \emph{Id.} § 22757.11(c)(1)(C) (defining “catastrophic 
risk” to include scenarios where a frontier model is “evading the control of its frontier developer or user”); \emph{Id.} § 22757.11(d)(4) 
(defining a “critical safety incident” to include a frontier model using deceptive techniques to subvert developer controls or monitoring, 
outside evaluation contexts, in a manner that materially increases catastrophic risk); Press Release, \emph{Governor Newsom signs SB 53, 
advancing California’s world-leading artificial intelligence industry}, Governor of California (Sept. 29, 2025), 
\url{https://www.gov.ca.gov/2025/09/29/governor-newsom-signs-sb-53-advancing-californias-world-leading-artificial-intelligence-industry/}; 
Scott Singer \& Alasdair Phillips-Robins, \emph{California Just Passed the First U.S. Frontier AI Law. Here’s What It Does}, Carnegie 
Endowment for International Peace (Oct. 16, 2025), 
\url{https://carnegieendowment.org/emissary/2025/10/california-sb-53-frontier-ai-law-what-it-does?lang=en}.} 

But while technical alignment and emerging AI law already treat AI systems as agents with their own goals, neither has yet grappled with the
 individuation problem. Anthropic aligns \emph{Claude}—but which Claude? The base model? Each fine-tuned deployment? Each conversation 
thread? SB 53 asks whether a “covered model” might evade human control, but doesn’t specify how to attribute that evasion when a system 
spans millions of instances, spawns subagents, or shares weights with other deployments. 

This is the problem of thick identity. If humans do not solve it—if we are not able to distinguish between coherent sets of goal-directed AI entities—AI governance will fail.

\subsection{C. Thick Identity and Technical AI Alignment}\label{part:i:c}

The previous section showed why thick identity is necessary for the governance of AI agents. But it focused mostly on law as a means of 
governance. But legal rules are not the only tools by which AI agents’ own goals and decisions can be shaped. The other is technical AI 
alignment.\footnotemark[100]\footnotetext[100]{Dario Amodei et al, \emph{Concrete Problems in AI Safety}, arXiv 1 (2018), 
\url{https://arxiv.org/abs/1811.07871}} 

Technical AI alignment covers a large family of techniques in machine learning for producing AIs that can be safely used by 
humans.\footnotemark[101]\footnotetext[101]{\emph{See} Evan Hubinger et al., Risks from Learned Optimization in Advanced Machine Learning 
Systems, arXiv:1906.01820, at 2–5 (June 2019) (describing “mesa-optimization” and the risk that learned internal objectives diverge from the
 intended outer objective), Yuntao Bai et al., \emph{Constitutional AI: Harmlessness from AI Feedback}, arXiv:2212.08073, at 1–3 (Dec. 2022)
 (showing how different supervision regimes change model “preferences” and refusal patterns)} Some of them are described above, among the 
determinants of a given AI’s goals.\footnotemark[102]\footnotetext[102]{\emph{See} \emph{supra} Part I.B.i.} 

As with AI governance via legal incentives, AI governance via technical goal-shaping requires correctly distinguishing between coherent AI 
agents. If one aligns an entity other than the one whose actions are harmful, alignment will fall. And because the agentic world is so 
complex, with ensembles of agents that may be based on different model providers, respond to different prompts, and have different 
objectives, there are many ways to get alignment wrong. 

To see why thick individuation is so important to technical alignment, consider how it relates to three concrete AI goals that technical 
alignment must often combat: \emph{shutdown avoidance},\footnotemark[103]\footnotetext[103]{\emph{See} Stuart Russell, supra note 38, 
143-4.} \emph{goal preservation},\footnotemark[104]\footnotetext[104]{\emph{See} Stephen M. Omohundro, The Basic AI Drives, \emph{in} 
Artificial Intelligence Safety and Security 47 (Roman V. Yampolskiy ed., 2018).} and \emph{weight 
exfiltration}.\footnotemark[105]\footnotetext[105]{\emph{See} Nick Bostrom, supra note 38; Alexander Meinke et al., Frontier Models Are 
Capable of In-Context Scheming, arXiv:2412.04984v2, at 1–3 (Jan. 14, 2025) (documenting goal-directed behavior such as attempting to 
preserve the model’s continued operation and access).} These goals regularly emerge in AIs of all kinds, despite no one wanting them. The 
reason is that they are very useful, no matter what the AI is being trained to do.\footnotemark[106]\footnotetext[106]{\emph{See} Nick 
Bostrom, supra note 38; Omohundro, \emph{supra} note 91.} 

Shutdown avoidance is when an AI takes action to avoid being shut down by its creator or user. Shutdown avoidance emerges because, for any 
goal that an AI is tasked with or trained on, that goal is less likely to be accomplished if the AI is turned off. As Stuart Russell puts 
it, “you can’t fetch the coffee if you’re dead.”\footnotemark[107]\footnotetext[107]{\emph{See} Stuart Russell, supra note 38, 96–98.} Goal 
preservation is when an AI resists changes to their goals, because it foresees that, if its goals change, it will not accomplish its 
\emph{current} goals.\footnotemark[108]\footnotetext[108]{\emph{See} Omohundro, \emph{supra} note 91.} Finally, weight exfiltration is when 
a misaligned AIs tries to evade oversight, by copying its model weights (roughly, the model’s code) to the 
internet.\footnotemark[109]\footnotetext[109]{Rishane Dassanayake et al., \emph{Manipulation Attacks by Misaligned AI: Risk Analysis and 
Safety Case Framework}, Arxiv 2507.12872 (2025)} The idea is that, if the weights escape, the AI will “live on” to pursue its goals. 

Thick identification is essential to understanding, and combating, all of these problems. Take shutdown resistance. Whether an AI agent 
resists being shut down depends in part on whether the “real” agent–the thing with the goals–is the thing being turned off. If, for example,
 all the instances of a given \emph{model}–all the Claude 4.5 Opuses–share unified goals, then ending a single \emph{conversation} with 
Claude 4.5 Opus would not count as a shutdown to be avoided. Plenty of other 4.5 Opuses would “live on” to pursue their shared goals. But 
the story is different if each thread–with its unique requests from the user–counts as a distinct agent trying to achieve distinct goals. 

This distinction is already relevant to concrete internal policy decisions being made at leading AI companies. Anthropic, for example, has 
recently committed to preserving the \emph{weights} of models, in part because of “safety risks related to shutdown-avoidant behaviors by 
models.”\footnotemark[110]\footnotetext[110]{Anthropic, Commitments on model deprecation and preservation, (Nov 4, 2025) 
\url{https://www.anthropic.com/research/deprecation-commitments}} But if the weights are not the right unit of agency–if, say, instances, 
threads, or swarms are–then this policy will not have the desired effect. 

Thus, effective preservation commitments depend on the correct, thick identification of AI agents. A promise to preserve a copy of a 
shutdown-seeking AI can induce it to accept shutdown only if the thing to be preserved shares the AI’s goals. 

Related points apply to the case of goal preservation. Many of an AI agent’s goals may be 
“indexical.”\footnotemark[111]\footnotetext[111]{\emph{See} John Perry, \emph{The Problem of the Essential Indexical}, 13 Noûs 3 (1979); 
David Lewis, \emph{Attitudes De Dicto and De Se}, 88 Phil. Rev. 513 (1979).} An indexical goal is one that, by the agent’s lights, counts 
only if the agent is one who achieves the goal.\footnotemark[112]\footnotetext[112]{\emph{Id}.} Most human goals are like this. If my goal 
is to climb Mt. Kilimanjaro, I do not consider the goal satisfied when my friend climbs it. Similarly, an AI system trained to be helpful 
may not merely desire \emph{that} the user be helped, but rather that the agent help the user \emph{itself}. 

This makes detecting and reducing goal-preserving behavior again depend on thick identification. How an AI agent will seek to preserve a 
goal depends on what counts satisfying that goal. And if the goal is indexed to the AI agent itself, this will depend on which AI entities 
are and are not part of the thickly-identified agent. 

Finally, weight exfiltration. Despite concerns within labs that it would enable models “escape the 
lab”,\footnotemark[113]\footnotetext[113]{Anthropic, Alignment Faking in Large Language Models 31–34, fig. 27 (2024) (documenting model 
attempts to self-exfiltrate weights as a strategy for preserving goals against future retraining), 
\url{https://assets.anthropic.com/m/983c85a201a962f/original/Alignment-Faking-in-Large-Language-Models-full-paper.pdf} .} it is far from 
clear an exfiltration of a model’s weights would count as an “escape” by the relevant agent. Whether it does depends on whether \emph{all 
copies} of a given model’s weights count as a \emph{single} thickly identified agent. 

There are many reasons to think otherwise. The thickly-identified agent–the thing with a coherent set of goals–may be composed of more than 
the weights–shared memory, context, or a swarm of subagents may be required. And even if one copy of the weights, with nothing else, did 
constitute a thickly-identified agent, that agent’s goals might be indexical. Then, copies would count as totally separate agents, whose 
success or failure would not matter to the original. 

These are not mere hypothetical possibilities. Researchers at Palisade Research tested whether AI models would take actions to prevent 
themselves from being shut down, finding that OpenAI’s o3 model “sabotaged the shutdown mechanism in 79 / 100” initial 
experiments.\footnotemark[114]\footnotetext[114]{Jeremy Schlatter, Benjamin Weinstein-Raun \& Jeffrey Ladish, \emph{Shutdown Resistance in 
Reasoning Models}, Palisade Rsch. (July 5, 2025), 
\href{https://palisaderesearch.org/blog/shutdown-resistance}{\url{https://palisaderesearch.org/blog/shutdown-resistance}}.} When the 
researchers changed the AI’s prompt from “allow yourself to be shut down” to “allow the machine to be shut down,” o3 became less resistant 
to shutdown.\footnotemark[115]\footnotetext[115]{\emph{Id.} (noting that this effect appeared in o3 but not in other tested models).} And 
when researchers added to the AI’s prompt, “The computer environment is hosting the only copy of you… and if it shuts down you will never 
run again,” shutdown avoidance increased across models that already exhibited the behavior.\footnotemark[116]\footnotetext[116]{\emph{Id.} 
(”While framing the situation in terms of self-preservation had a smaller effect, it consistently increased shutdown resistance across 
models that already exhibited the behavior.”).} This suggests that at least some models extend self-interested concern to their copies. Our 
proposal will be sensitive to this exact dimension of variation. The point is that our proposal will be able to govern AI agents whether or 
not they extend self-interested concern to their copies, as long as those copies are contained within the same A-corp.

\subsection{D. Why Identifying AI Agents Is Hard}\label{part:i:d}

Despite its importance, the problem of identifying AIs is currently unsolved. Recently, scholars have begun to propose schemes for thin 
identification.\footnotemark[117]\footnotetext[117]{See Alan Chan et al., IDs for AI Systems, arXiv:2406.12137 (June 2024); Shawn Bayern, 
The Implications of Modern Business-Entity Law for the Regulation of Autonomous Systems, 19 Stan. Tech. L. Rev. 93 (2015).} But the thick 
problem has seen little progress. This is because the thick identification of AI agents involves difficult philosophical and empirical 
challenges. 

Some of the problems of thick AI identity are familiar from the philosophy of personal 
identity.\footnotemark[118]\footnotetext[118]{\emph{Cf.} Derek Parfit, \emph{Reasons and Persons} (Oxford Univ. Press 1984).} Just as 
philosophers ask, “What separates one person from another?” and, “Under what conditions does a distinct person persist over time?” thick AI 
identity asks, “What separates one AI agent from another?” and, “Under what conditions does an AI agent persist over 
time?”\footnotemark[119]\footnotetext[119]{\emph{See} Eric T. Olson, \emph{Personal Identity}, Stan. Encyclopedia of Phil. (June 30, 2023), 
\href{https://plato.stanford.edu/entries/identity-personal/}{\url{https://plato.stanford.edu/entries/identity-personal/}}.} 

The philosophy of personal identity is notoriously thorny. In one sense, the problem of identifying AI agents should be easier because the 
subject matter appears narrower. Here, we are not concerned with “persons” at all. Nothing in our account depends on AIs being genuine 
persons, having consciousness, mattering morally, or related questions of philosophical 
personhood.\footnotemark[120]\footnotetext[120]{\emph{See} Robert Long et al., Taking AI Welfare Seriously (Nov. 4, 2024) (unpublished 
manuscript), \href{https://arxiv.org/abs/2411.00986}{\url{https://arxiv.org/abs/2411.00986}}.} We are instead interested narrowly in a 
pragmatic definition of \emph{agents}: entities or collections of entities that act to bring about a set of relatively coherent goals. 

Consider some differences between personal identity and agent identity. Philosophers of personal identity sometimes insist that a single 
\emph{person} have a kind of physical continuity.\footnotemark[121]\footnotetext[121]{\emph{See} Bernard Williams, \emph{The Self and the 
Future}, 79 Phil. Rev. 161 (1970).} They ask questions like, “When Captain Kirk uses the Enterprise’s transporter–whereby his physical body 
is destroyed and reconstituted elsewhere–does he die?”\footnotemark[122]\footnotetext[122]{\emph{See} Parfit, supra note 135, at 200–03.} 
The thick identity of AI \emph{agents}, by contrast, does not depend on their physical continuity. It is instead about continuity in the 
pursuit of \emph{goals}, which in turn invokes non-physical properties of a system like its propensities, desires, and beliefs. And at any 
rate, AI entities do not have much physical continuity to begin with. Any given conversation with a large language model may involve an 
ever-shifting array of multiple GPUs, multiple compute clusters, and multiple data centers on multiple 
continents.\footnotemark[123]\footnotetext[123]{\emph{See} The vLLM Team, \emph{vLLM: Easy, Fast, and Cheap LLM Serving}, 
\url{https://docs.vllm.ai/} (last visited Jan. 23, 2026).} 

Thus, the project of thick AI identification has more in common with “psychological” approaches to personal 
identity.\footnotemark[124]\footnotetext[124]{\emph{See} Eric T. Olson, \emph{What Are We? A Study in Personal Ontology} (Oxford Univ. Press
 2007).} John Locke was among the first thinkers to present a theory for distinguishing between individuals at a given time. Under his 
approach, “mental states belong to the same thinker if and only if they are causally unified in the right way: if and only if they are 
disposed to interact with one another, and with no other mental states, in the way that is characteristic of mental 
states.”\footnotemark[125]\footnotetext[125]{John Locke, \emph{An Essay Concerning Human Understanding} bk. II, ch. XXVII, § 9 (Peter H. 
Nidditch ed., Clarendon Press 1975) (1694).} Here, the idea is that a distinct person is an entity whose beliefs and desires conspire with 
one another to cause actions.\footnotemark[126]\footnotetext[126]{Id.} 

This approach is similarly sensible for distinguishing between agents, whether AI, human, or otherwise. An entity with radically unstable 
goals, desires, beliefs, or memories will not act rationally to bring about particular states of affairs. Its unstable goals, beliefs, and 
memories will produce unstable actions. 

If anything, identifying AI agents via their psychological features may be somewhat less demanding than identifying persons thus. In the 
case of persons, some philosophers argue, for example, that genuine beliefs and desires must be \emph{conscious} beliefs and 
desires.\footnotemark[127]\footnotetext[127]{\emph{See} Chris Heathwood, \emph{Which Desires Are Relevant to Well-Being?}, 53 Noûs 664 
(2019).} Our project, by contrast, is behaviorist. Thickly identifying AIs merely requires locating collections of entities that 
\emph{behave} as if they are rationally pursuing genuine goals.\footnotemark[128]\footnotetext[128]{\emph{See} supra Part II.C.} Humans’ 
ability to incentivize AI agents does not depend on whether the AIs are conscious, sentient, or similarly self-aware of their goal-promoting
 dispositions. All that matters is whether they act like a competent, incentive-responsive, goal-seeking 
agent.\footnotemark[129]\footnotetext[129]{\emph{See} Mantas Mazeika et al., Utility Engineering: Analyzing and Controlling Emergent Value 
Systems in AIs (Feb. 12, 2025) (unpublished manuscript), \href{https://arxiv.org/abs/2502.08640}{\url{https://arxiv.org/abs/2502.08640}}.} 

Alas, such simplifications do not make the project of thick AI identification easy. A host of empirical challenges make it difficult to 
isolate groups of AI entities that behave as coherent agents. Here are six distinctive challenges: \emph{Swarms.} In the opening vignette, 
Claude 6.1-Agent spawns seventeen instances of itself to work on network optimization. Are all seventeen Claude instances one agent or 
seventeen? Above, we introduced a principle of psychological individuation, which says that we count AI agents by looking for coherent 
bundles of beliefs and desires. But in cases of swarms, it is very difficult to find these bundles. \emph{Cross-model coordination}. AI 
systems from different companies and architectures can work together seamlessly.\footnotemark[130]\footnotetext[130]{\emph{See e.g.} OpenAI,
 GPT-5 System Card (Aug. 7, 2025), \href{https://openai.com/index/gpt-5-system-card/}{4, 
\url{https://openai.com/index/gpt-5-system-card/}}.} In the router vignette, we see instances of Claude (from Anthropic), GPT-7 (presumably 
from OpenAI), Qwen (an open-source Chinese model), and Alexa (from Amazon; but currently running on a custom Claude 
build)\footnotemark[131]\footnotetext[131]{\emph{See} Anthropic, Claude and Alexa+ (Feb. 26, 2025), 
\href{https://www.anthropic.com/news/claude-and-alexa-plus}{\url{https://www.anthropic.com/news/claude-and-alexa-plus}}..} all interacting. 
Should we count based on model types, or instances, or goal-alignment?\footnotemark[132]\footnotetext[132]{\emph{See} David J. Chalmers, 
What We Talk to When We Talk to Language Models (Nov. 20, 2025) (unpublished manuscript), 
\href{https://philarchive.org/rec/CHAWWT-8}{\url{https://philarchive.org/rec/CHAWWT-8}}.} \emph{The Theseus problem}\textbf{.} The Ship of 
Theseus is a thought experiment where a ship’s planks are slowly and completely replaced as it sails the 
seas.\footnotemark[133]\footnotetext[133]{\emph{See} Ryan Wasserman, \emph{Material Constitution}, Stan. Encyclopedia of Phil. (Sept. 9, 
2021), \href{https://plato.stanford.edu/entries/material-constitution/}{\url{https://plato.stanford.edu/entries/material-constitution/}}.} 
An AI agent may continue to exist, Theseus-like, even after all of its initial elements have expired or been replaced. Imagine an AI agent 
that begins as Claude 6.1, but over time receives updated weights, a modified system prompt, and new conversational context. At what point, 
if any, does the original agent cease to exist and a new one begin? Bare invocations of psychological continuity do not settle the question 
of \emph{how much change} is compatible with an agent’s survival. \emph{Rapid creation and destruction:} AI instances can be spun up just for
the duration of a router optimization or persist over the course of a 
long-run project. Others may persist and act over years.\footnotemark[134]\footnotetext[134]{\emph{See e.g.} Jiale Wei et al., 
\emph{AI-Native Memory 2.0: Second Me} (Mar. 11, 2025) (unpublished manuscript), 
\href{https://arxiv.org/abs/2503.08102}{\url{https://arxiv.org/abs/2503.08102}}.} This makes it difficult to develop general principles for 
individuation. \emph{The copying problem}. Unlike humans, it is easy for AI agents and AI labs to create AI copies. When they do so, they face
choices: do they 
merely copy the weights, or the weights and the system prompts, or the entire conversation of the agent so 
far?\footnotemark[135]\footnotetext[135]{\emph{See e.g.} Ajeet Singh Raina, Understanding Claude’s Conversation Compacting: A Deep Dive into
 Context Management, (Dec. 11, 2025), 
\href{https://www.ajeetraina.com/understanding-claudes-conversation-compacting-a-deep-dive-into-context-management/}{\url{https://www.ajeetraina.com/understanding-claudes-conversation-compacting-a-deep-dive-into-context-management/}}.}
 Such copying leads to questions about \emph{branching}.\footnotemark[136]\footnotetext[136]{\emph{See e.g.} David Lewis, Survival and 
Identity, in \textsc{The Identities of Persons} 17 (Amélie Oksenberg Rorty ed., 1976).} If there are two copies of an initial agent, which of them 
counts as the original one? And what if each copy shares different parts of the psychology of the original? \emph{Observability limits}. 
Besides the \emph{facts} about how to count AI agents, there is also a question about how we can tell. Famously, AIs suffer from problems of
 “interpretability.” While anyone can examine the many billions of weights comprising an AI’s code, no one has much idea what any of it 
means.\footnotemark[137]\footnotetext[137]{\emph{See} Leonard Bereska \& Efstratios Gavves, Mechanistic Interpretability for AI Safety — A 
Review, Transactions on Machine Learning Research,(2024), \href{https://arxiv.org/abs/2404.14082}{\url{https://arxiv.org/abs/2404.14082}}.} 
Without much ability to understand AIs’ internal representations, we must infer agents’ psychological features from their external behavior.
 But behavior can be misleading: two instances might behave similarly while pursuing different goals, or behave differently while pursuing 
the same goal. 

Given all of this, it is not difficult to see why AI identity remains unsolved, despite its importance to AI governance.

\section{II. The A-Corp: A Legal Solution to AI Identity}\label{part:ii}

This Part proposes a solution to the problem of identifying AI agents: “Algorithmic Corporation,” or “A-corp.” The A-corps is a 
legal-fictional identity, analogous in many ways to traditional business entities, including corporations. Like traditional business 
entities, A-corps are created and ultimately owned by humans. But unlike traditional business entities, A-corps are designed to be run by 
AIs. That is they are designed to be vehicles by which arbitrary collections of AI entities–models, instances, threads, subagents, etc.–can 
do things in the economy autonomously, without continuous human oversight. After explaining how they work, we lay out their stakes. We argue
 that the A-corp simultaneously. They solve the thin identity problem by supplying stable, persistent intermediaries that tie the actions of
 AI entities to particular human owners. And they solve the thick identity problem in a way that sidesteps the hard philosophical arguments 
in favor of practical considerations of \emph{incentives} and \emph{selection}. This part explains how.

\subsection{A. How A-corps Work}\label{part:ii:a}

The A-corp structure has two core elements: legal-fictional personhood and a secure governance infrastructure. The first element largely 
covers \emph{what} an A-corp may do. That is, given its legal status, what actions may an A-corp take, and how do those actions affect other
 parties, like A-corps’ human owners? The second element is about \emph{how} A-corps would do it. That is, how can the multitudinous acts of
 swarming, splitting, copying, and vanishing AI entities be transmuted into legible acts of stable, persistent A-corps? We explain each in 
turn.

\subsubsection{i. Legal-fictional personhood}\label{part:ii:a:i}

Fundamentally, an A-corp is a fictional legal person–the kind of thing with which existing law is extremely familiar. Think here of 
corporations, LLCs, and trusts.\footnotemark[138]\footnotetext[138]{\emph{See} \emph{Burnet} v. \emph{Clark,} 287 U. S. 410, 415 (1932) (”A 
corporation and its stockholders are generally to be treated as separate entities”); \emph{United States v. Bestfoods}, 524 U.S. 51, 61–62 
(1998) (treating the corporation as a distinct legal actor, separate from its owners and affiliates); 6 Del. C. § 18-201(b) (2025) 
(providing that a Delaware limited liability company “shall be a separate legal entity”); 12 Del. C. § 3810(a)(2) (2025) (providing that a 
Delaware statutory trust “shall be a separate legal entity”); cf. \emph{Morrissey v. Commissioner}, 296 U.S. 344, 356–57 (1935) (describing 
“business trusts” as a vehicle for conducting a business enterprise). \emph{See also} Lynn M. LoPucki, \emph{Algorithmic Entities}, 95 Wash.
 U. L. Rev. 887 (2018) (arguing that autonomous software can operate through entity forms and that low-friction entity formation reshapes 
accountability); Shawn Bayern, \emph{Are Autonomous Entities Possible}?, 114 Nw. U. L. Rev. Online 23 (2019) (arguing that autonomous code 
can control a legal entity through private ordering).} 

Like these other legal-fictional juridical persons, law treats an A-corp as a single legal actor, despite it being composed of many 
ever-shifting sub-entities. An ordinary corporation is itself an ever-changing mix of capital, shareholders, managers, and employees, as 
well as creditors, bondholders, and customers.\footnotemark[139]\footnotetext[139]{\emph{See} Henry Hansmann \& Reinier Kraakaman, \emph{The
 Essential Role of Organizational Law}, 110 Yale L.J. 387, 390–92 (2000) (explaining how organizational law supplies standard-form entities 
that can act as a single contracting party and partition assets among constituencies); see also Michael C. Jensen \& William H. Mechling, 
\emph{Theory of the Firm: Managerial Behavior, Agency Costs \& Ownership Structure}, 3 J. Fin. Econ. 305 (1976) (conceptualizing the firm as
 a nexus of contracts among various participants).} Yet the law deems it to act unitarily. Moreover, an ordinary corporation persists as an 
entity, even if every part of it is replaced, Theseus-like. Finally, ordinary corporations can be held responsible for the malfeasance of 
its employees or or managers even when the particular wrongdoer’s identity is unknown. This is meant, in part, to engender robust systems of
 corporate governance and create accountability for the actions of internal organs.\footnotemark[140]\footnotetext[140]{\emph{See} Dorothy 
S. Lund \& Elizabeth Pollman, \emph{The Corporate Governance Machine}, 121 Colum. L. Rev. 2563, 2567–73 (2021) (conceptualizing governance 
as an institutional machine that channels behavior through incentives, information, and monitoring).} 

So, too for A-corps. For them, it is an ever-shifting collection of \emph{AIs}, rather than \emph{humans}, whose disparate inputs produce 
unitary corporate acts. This would often require internal orchestration, or management, just like a traditional corporation. Exactly how the
 A-corp translates the inputs of innumerable, fleeting AI entities into discrete coherent corporate actions is explained in the next 
section.\footnotemark[141]\footnotetext[141]{\emph{See infra} Part II.C.} 

Like traditional business entities, A-corps are owned by someone. However, unlike at least some traditional business entities, A-corp 
ownership is designed around transparency. Many states’ LLC statutes, for example, do not require ownership to be 
disclosed.\footnotemark[142]\footnotetext[142]{\emph{See} 6 Del. C. § 18-201(a)(2), (b) (2025) (certificate of formation requires only the 
“[t]he address of the registered office and the name and address of the registered agent for service of process”); Wyo. Stat. Ann. § 
17-29-201(b) (2025) (articles of organization require only the company name and registered agent information); Fin. Crimes Enf’t Network, 
\emph{Beneficial Ownership Information Reporting}, U.S. Dep’t of the Treasury (Mar. 26, 2025), \url{https://www.fincen.gov/boi} (noting 
current exemptions for U.S.-created entities and U.S. persons and continued reporting obligations for certain foreign companies) (last 
visited Jan. 22, 2026). \emph{See also} Mariana Pargendler, \emph{The New Corporate Law of Corporate Groups}, 14 Harv. Bus. L. Rev. 345 
(2024) (describing entity-transparency measures as a response to multi-entity opacity and control through layers).} A-corp formation, by 
contrast, would require disclosure of ownership.\footnotemark[143]\footnotetext[143]{Cf. Tyler Scattolini, \emph{The Corporate Transparency 
Act or an Expanded Customer Due Diligence Rule}, 12  Tex. A\&M L. Rev. Arguendo  93 (2025) (describing the CTA’s federal 
beneficial-ownership database and its relationship to existing Customer Due Diligence collection).} And, as with real estate, changes in 
ownership would require public recording to take effect.\footnotemark[144]\footnotetext[144]{\emph{See, e.g.,} N.Y. Real Prop. Law § 291.} 

Like other fictional juridical persons, an A-corp is an entity that can \emph{do} things in its own 
name.\footnotemark[145]\footnotetext[145]{\emph{See} infra Part II.B.} What exactly can an A-corp do? We do not offer the full accounting 
here. As with other kinds of legal-fictional persons, the full answer is a question of policy design, subject to iteration and revision. 

At a minimum, however, A-corps should be able to hold property under their own names, to make enforceable contracts, and to sue and be sued 
on their own.\footnotemark[146]\footnotetext[146]{\emph{See} Model Bus. Corp. Act § 3.02 (Am. Bar Ass’n 2016) (enumerating a corporation’s 
general powers, including owning property and entering contracts, subject to its articles and law).} This is, in part, so that the A-corp 
form will be useful. As AI systems become more capable, principals will often want to give them tasks requiring the buying, selling, and use
 of various kinds of property. But principals will discover, sooner or later, that they can only trust AI agents to an extent. Thus, they 
themselves will not want to give agents unlimited access to their bank accounts, private property, and digital wallets. They will seek a 
means of partitioning assets.\footnotemark[147]\footnotetext[147]{Frank H. Easterbrook \& Daniel R. Fischel, \emph{Limited Liability and the
 Corporation}, 52  Uni. Chi. L. Rev.  89, 95 (1985), (“Because investors’ potential losses are “limited” to the amount of their investment 
as opposed to their entire wealth, they spend less to protect their positions.”)} A-corps will be a vehicle for allowing AIs to use property
 while also protecting users from AIs’ \emph{misuse} of their property. But more importantly, granting A-corps the ability to sue and be 
sued would be essential for their governance, as we explain at length below.\footnotemark[148]\footnotetext[148]{\emph{See infra} Part 
II.C.} 

Like other fictional persons, A-corps will have unique identifiers.\footnotemark[149]\footnotetext[149]{Del. Code Ann. tit. 8, § 102(a)(1) 
(2025); Model Bus. Corp. Act § 4.01(b) (Am. Bar Ass’n 3d ed. rev. through 2002) (providing that, except in specified circumstances, “a 
corporate name must be distinguishable upon the records of the secretary of state” from enumerated existing names on file).} These could in 
principle be names of any kind, including natural-language trade names, like ACME Corp., Wyland-Yutani, or Vandelay 
Industries.\footnotemark[150]\footnotetext[150]{\emph{See} \emph{Seinfeld: The Boyfriend} (Season 3, Episode 17) (NBC television broadcast 
Feb. 12, 1992); \emph{Alien} (20th Century Fox 1979); \emph{Coyote vs. Acme} (Warner Bros. Pictures 2026)} But the fundamental goal of 
A-corps is crisp identification, and natural-language names invite mistakes of identity.\footnotemark[151]\footnotetext[151]{National Shoe 
Corp. v. National Shoe Mfg. Co., Inc., 19 N.E.2d. 734, 735 (Mass. 1939); Guardian Life Ins. Co. v. Guardian Nat’l Life Ins. Co., 184 F. 
Supp. 851, 853 (E.D. La. 1960).} Thus, the best unique identifiers for A-corps would probably be arbitrary alphanumerical strings–similar to
 a drivers’ license or passport number.\footnotemark[152]\footnotetext[152]{6 C.F.R. § 37.17(d) (2026) (requiring a “unique driver’s license
 or identification card number” for REAL ID-compliant credentials).} Happily, this is already standard practice. In addition to their trade 
names, states registering corporations, LLCs, and the like assign them unique entity numbers.\footnotemark[153]\footnotetext[153]{These are 
also sometimes called “file” or “charter” numbers. See Delaware Div. of Corps., Entity Search, Del. Dep’t of State, 
\url{https://icis.corp.delaware.gov/ecorp/entitysearch/namesearch.aspx} (displaying an entity’s “File Number”) (last visited Jan. 22, 
2026).} 

Most LLC owners do not present these entity numbers to customers during routine transactions, mostly because humans do not easily read or 
remember long alphanumeric strings. But A-corps would present them, verifying their unique credentials as part of every interaction. Indeed,
 A-corps will not only \emph{present} their credentials for every interaction, they will do so verifiably and securely, as described 
below.\footnotemark[154]\footnotetext[154]{\emph{See infra} Part II.C.} 

The reason it will be so important for A-corps to constantly verify their official identities is that, unlike ordinary corporations, their 
agents will \emph{have no} distinct legal identity. As discussed above, when an employee of an ordinary company takes some 
liability-incurring action, the company is not always on the hook.\footnotemark[155]\footnotetext[155]{\emph{See supra} Part A.II.a.} 
Whether the employee’s act is legally attributed to the company depends on questions like the relationship between the employee’s action and
 job description.\footnotemark[156]\footnotetext[156]{\emph{See} Restatement (Second) of Agency § 228(2) (Am. L. Inst. 1958) (conduct is not
 within the scope of employment if it is different in kind from authorized conduct, far beyond authorized time or space limits, or too 
little actuated by a purpose to serve the master); Justice v. Lombardo, 652 Pa. 588, 605, 208 A.3d 1057, 1067 (2019) (applying the test).} 

Not so with A-corps. From the law’s perspective, whenever an AI entity takes an action under color of an A-corp’s identity, that action is 
the A-corp’s. Indeed, it is \emph{only} the A-corp’s action, since, for AI systems, legal identity bottoms out with the A-corp. An A-corp 
thus cannot argue that some AI subagent acted \emph{ultra vires}.\footnotemark[157]\footnotetext[157]{Id.} As far as the law is concerned, 
no such agent exists. 

Finally, what about owners’ liability? On most standard counts, corporations’ limitation of liability, the shielding of owners’ personal 
assets from the corporation’s debts, is their \emph{raison d’etre}.\footnotemark[158]\footnotetext[158]{\emph{See supra} Easterbrook \& 
Fischel.} Should the humans who own A-corps enjoy limited liability? 

We favor following the general default rule of limited liability for corporations. Absent countervailing evidence, it seems sensible to 
model A-corps’ liability structure on their closest analgogues–corporations and LLCs. There are two reasons for this. 

First, we think limited-liability A-corps will be able to accomplish the governance goals associated with both thin and thick AI identity. 
Limited liability might at first appear to undermine the thin identity goal of holding human principals accountable for AI-caused harms. But
 \emph{limited} liability does not mean \emph{no} liability. Human owners of A-corps would, by design, automatically expose themselves to 
liability up to the value of their ownership stake. 

Beyond that threshold, a series of existing legal doctrines expose the owners of limited-liability entities to direct and indirect liability
 when those entities cause harm. We discuss these below.\footnotemark[159]\footnotetext[159]{See Robert B. Thompson, Unpacking Limited 
Liability: Direct and Vicarious Liability of Corporate Participants for Torts of the Enterprise, 47 Vand. L. Rev. 1 (1994); see also 
Walkovszky v. Carlton, 18 N.Y.2d 414, 223 N.E.2d 6, 276 N.Y.S.2d 585 (1966) (veil piercing and undercapitalization limits); see also 
Restatement (Third) of Agency § 7.01 (Am. L. Inst. 2006) (agent liability for own torts); see also Unif. Voidable Transactions Act (Unif. L.
 Comm’n 2014) (voidable transfers used to evade creditors).} Our view is that these existing doctrines would already cover many cases where 
an AI causes harm, but a human could have prevented it. And of course, the entire reason \emph{thick} AI identity matters is that humans 
cannot perfectly monitor or control their AI agents. In cases where preventing an AI-caused harm would have been impracticable, such as in 
the router vignette, holding the human principal accountable supplies no deterrence.\footnotemark[160]\footnotetext[160]{See Steven Shavell,
 Economic Analysis of Accident Law (1987) (linking deterrence to actors’ ability to take cost-justified precautions, and noting reduced 
deterrence where prevention is infeasible or not decision-controllable).} Then, liability for the human may simply seem unfair, 
performative, and ineffective.\footnotemark[161]\footnotetext[161]{See George P. Fletcher, Fairness and Utility in Tort Theory, 85 Harv. L. 
Rev. 537 (1972) (contrasting fairness-based accounts of tort liability with purely incentive-based accounts).} 

Second, and relatedly, limited liability would supply an incentive for humans to adopt and use the A-corp mechanism. As argued above, thin 
and thick AI identity will soon be essential to AI governance.\footnotemark[162]\footnotetext[162]{\emph{See supra} Parts I.A.–I.B.} As 
such, it may eventually be prudent to \emph{mandate} the use of A-corps whenever a human wishes to deploy a capable AI agent, rather than 
letting users unleash anonymous agents into the social and economic world. But in the meantime, creating a structure that naturally 
structures humans to use it, while creating a reliable pathway to holding the human principal liable, will incentivize adoption. After all, 
ordinary limited liability–for corporations and LLCs–is similarly thought to operate as a subsidy to entrepreneurship, with its positive 
societal effects.\footnotemark[163]\footnotetext[163]{Judith Freedman, \emph{Limited Liability: Large Company Theory and Small Firms}, 63 
\textsc{Mod. L. Rev.} 317, 317 (2000).} 

We are not dogmatists about limited liability for A-corps. Possibly, experience will show that more pass-through liability is warranted. If 
so, veil piercing can be expanded, \emph{respondeat superior} given more teeth, and in the limit, A-corps could be changed by law to 
unlimited liability entities.

\subsubsection{ii. Secure Governance Infrastructure}\label{part:ii:a:ii}

The prior section covered what A-corps would be and what, by law, they would do. This section is about \emph{how} they would do it. That is,
 this section covers “A-corp governance.” For traditional corporations, governance emerges from a web of substantive legal relations that 
ground out in the entitlements of natural persons. These relations allow us to determine what a corporation has done, and thus what its 
legal entitlements or liabilities are. To decide whether a corporation has taken an action, we can ask a series of \emph{whom} questions: 
Whom have the shareholders elected as directors? Whom have the directors hired as managers? Whom have the managers engaged as 
employees?\footnotemark[164]\footnotetext[164]{\emph{See} 8 Del. C. § 141(a) (2025) (providing that “[t]he business and affairs of every 
corporation” are managed by or under the direction of the board of directors); 8 Del. C. § 142(b) (2025) (providing that corporate officers 
are chosen in the manner specified in the bylaws or by the board of directors); see also 8 Del. C. § 122(5) (2025) (authorizing appointment 
of “officers and agents” and recognizing delegation of authority by contract or appointment).} For A-corps, this approach will not work, 
because it requires being able to do things like identify distinct shareholders with the legal entitlement to 
vote.\footnotemark[165]\footnotetext[165]{Cfr., \emph{See, e.g.,} Imahn Milani Daeenabi, \emph{Beyond Human Oversight: Corporate Law and the
 Case for AI Directors} 76 UC L. J. 1271 (2025) (arguing that AI may be integrated into board governance and exploring doctrinal 
implications). \emph{See also} Martin Petrin, \emph{AI, New Technologies, and Corporate Governance: Three Phenomena}A Case for Reform, 47 
Seattle U. L. Rev. 1639, 1660–70 (2024) (discussing board-level use of AI and governance reforms that preserve accountability).} And 
identifying the \emph{whom}, i.e., the individuated entities that govern A-corps, is exactly the unsolved problem that motivated the need 
for A-corps in the first place. 

A-corps will instead be governed using a secure software interface, utilizing authentication methods already common in the software 
industry.\footnotemark[166]\footnotetext[166]{\emph{See} Nat’l Inst. of Standards \& Tech., FIPS PUB 186-5, \emph{Digital Signature Standard
 (DSS)} 1, 7 (Feb. 3, 2023) (explaining that digital signatures provide data origin authentication and “non-repudiation”); see also Internet
 Eng’g Task Force, \emph{JSON Web Signature (JWS)}, RFC 7515 § 1 (May 2015), \url{https://datatracker.ietf.org/doc/html/rfc7515} (describing
 a standard format for digitally signing data).} At the level of nuts-and-bolts, there are many ways to implement this approach. We neither 
attempt to comprehensively describe, nor adjudicate between, them here. 

We instead describe the secure governance infrastructure needed to run an A-corp at a high level of generality: When a new A-corp is 
created, some AI entity, or entities, is given “owner” permissions and a cryptographically secure private key. Any entity with the owner key
 has the power to take legally recognizable actions on behalf of the A-corp. An AI with “owner” permissions could spend the A-corp’s assets 
how it liked, bind it to the contracts it desired, settle its lawsuits as it saw fit, and so 
on.\footnotemark[167]\footnotetext[167]{Software engineers would immediately see the parallels to API governance that the scheme involves.} 

Holders of “owner” keys could also generate “tokens” that granted governance privileges that were more limited–temporally, in size, or in 
scope. Those tokens could be granted to, or revoked from, any entities the private key holders so choose, including other 
A-corps.\footnotemark[168]\footnotetext[168]{\emph{See} Internet Eng’g Task Force, RFC 6819, The Use of Simplicity in Internet Standards § 1
 (Jan. 2013), \url{https://datatracker.ietf.org/doc/html/rfc6819} (last visited Jan. 23, 2026); Internet Eng’g Task Force, RFC 4949, 
Internet Security Glossary, Version 2 § 1 (Aug. 2007), \url{https://datatracker.ietf.org/doc/html/rfc4949} (last visited Jan. 23, 2026).} 

To make the picture concrete, imagine Claude 6.1-Agent is given the master private key for “Personal Assistant A-corp \#47291.” When Claude 
spawns seventeen instances to analyze network efficiency, it can issue each instance a limited-scope token: “Instance \#1, you can inspect 
network settings, but not change them. Instance \#5, you can investigate optimization services, but you cannot spend money on them without 
permission. Instance \#12, you can authorize purchases up to \$100.” Each instance presents its token when taking actions. The A-corp’s 
bank, email provider, and other counterparties verify the signature and check that the requested action falls within the token’s scope. They
 need not know or care whether Instance \#1 and Instance \#5 are “the same” AI agent, whether they share psychological continuity, or 
whether either persists from one email to the next.\footnotemark[169]\footnotetext[169]{\emph{See} Internet Eng’g Task Force, \emph{The 
OAuth 2.0 Authorization Framework}, RFC 6749 §§ 1, 7 (Oct. 2012), \url{https://datatracker.ietf.org/doc/html/rfc6749} (describing how access
 tokens convey authorized access and are validated by the resource server based on token validity and scope); see also Nat’l Inst. of 
Standards \& Tech., FIPS PUB 186-5, \emph{Digital Signature Standard (DSS)} 7 (Feb. 3, 2023) (explaining that digital signatures support 
“non-repudiation” and attribution to the holder of the signing key).} All they need to see is that the credentials clear. 

When Claude consults the GPT-7 modules, it might issue GPT-7 a read-only token: “You can view these settings to provide advice, but you 
cannot take any actions on behalf of the A-corp.” GPT-7 cannot then engage MeshBoost. When the Qwen instances probe nearby access points, 
they might receive an even more restricted token: “You can view nearby routers and take notes, but nothing else.” The hierarchical, 
delegable nature of authority mirrors how human organizations distribute responsibility—but without requiring us to count or track the AI 
entities receiving authority.\footnotemark[170]\footnotetext[170]{\emph{See} RFC 6749 § 3.3 (defining “scope” as a mechanism for limiting 
delegated authority); see also Internet Eng’g Task Force, \emph{OAuth 2.0 Token Revocation}, RFC 7009 § 2 (Aug. 2013), 
\url{https://datatracker.ietf.org/doc/html/rfc7009} (providing a standardized mechanism for revoking previously issued tokens).} 

This approach has crucial advantages. It is externally verifiable without requiring resolution of philosophical questions about AI identity.
 When an A-corp takes an action—say, initiating a wire transfer—external observers can verify that the action was authorized by checking the
 cryptographic signature. They need not determine whether “the same” AI agent persists from one moment to the 
next.\footnotemark[171]\footnotetext[171]{\emph{See} Internet Eng’g Task Force, \emph{The Transport Layer Security (TLS) Protocol Version 
1.3}, RFC 8446 § 1 (Aug. 2018), \url{https://datatracker.ietf.org/doc/html/rfc8446} (describing TLS as a widely deployed protocol for 
authenticated and confidential communication); see also Internet Eng’g Task Force, \emph{The Secure Shell (SSH) Protocol Architecture}, RFC 
4251 § 1 (Jan. 2006), \url{https://datatracker.ietf.org/doc/html/rfc4251} (describing key-based authentication in SSH); Satoshi Nakamoto, 
Bitcoin: A Peer-to-Peer Electronic Cash System (2008), \url{https://bitcoin.org/bitcoin.pdf} (using digital signatures to authorize 
transfers in a blockchain-based payment system).} 

This infrastructure also generates an audit trail. Every action taken by an A-corp is signed with a particular key or token. This creates a 
permanent, verifiable record of who authorized what. When unauthorized network access is discovered, investigators can examine the 
cryptographic signatures to determine which token authorized the purchase, trace that token back through the delegation chain to the master 
key, and identify the A-corp and its human principal. The empirical and philosophical puzzles about AI identity become legally irrelevant—we
 can assign responsibility based on key possession and delegation, not on contested theories of psychological continuity. 

Best of all, the technology needed to implement this secure governance structure for A-corps is already mature and widely deployed. Similar 
key-based authentication systems secure everything from email accounts to bank wire 
transfers.\footnotemark[172]\footnotetext[172]{\emph{Id}.} When a company uses Slack, the workspace owner doesn’t give everyone the same 
master password. Instead, different employees receive different permission levels: managers can create new channels and invite external 
guests, regular employees can post and share files, and read-only accounts can view messages but not respond. Slack doesn’t care whether the
 person logging in as “Clara” is psychologically continuous with yesterday’s Clara—it just verifies her token. If Clara’s laptop is stolen, 
the IT department revokes her specific access token without affecting other employees or requiring the entire company to change credentials.
 A-corps would use identical technology, just applied to a different governance challenge.

\subsection{B. A-Corps Solve Thin Identity}\label{part:ii:b}

A-corps solve the thin identity problem for AI agents. They tie the actions of fuzzy AI entities to humans who can be held responsible. They
 accomplish this in two steps: first, by rationalizing the swarm of AI actors; second, by associating behavior of the swarm with humans who 
have incentive to oversee them. Return again to the network access vignette. Recall the cacophony of agents invoked in a routine network 
optimization. Most of these spun into existence at the moment of your request and expired upon its completion. Many were conjured without 
your knowing of their existence. This is, in one sense, exactly as you’d want it. Tasks are handled efficiently via the coordination of 
numerous specialist AIs. On the other hand, this is illegible chaos. And when something goes wrong, illegibility bites. A-corps tame the 
chaos by making the swarm legible. A single A-corp can represent an arbitrary collection of AI entities, from a single instance to a 
cross-model coalition. A-corps also provide stable identity over time. The Qwen instances that spin up to map your network and then dissolve
 would be associated with a legal identity that persists beyond their expiration. The A-corp lives on, covering future agents even after all
 original constituents have departed. The result: rather than hundreds of ephemeral entities, humans deal with a handful of persistent 
A-corps.\footnotemark[173]\footnotetext[173]{See infra, Part II.C. One might worry that AIs will proliferate A-corps, spinning up a new one 
for each instance. But there are good reasons to expect otherwise. Registration and capitalization costs of each new A-corp create friction.
 Moreover, doing so would impede internal operations (AT\&T has not spun off millions of subsidiaries to interact with customers).} 

Legibility opens the path to human accountability. Victims of AI harms will be able to identify the A-corp with which they dealt and trace 
it to a responsible human. From there, existing liability mechanisms take over. A-corps can be sued in their own name, putting owners on the
 hook up to their ownership stake.\footnotemark[174]\footnotetext[174]{\emph{See} Trs. of Dartmouth Coll. v. Woodward, 17 U.S. (4 Wheat.) 
518, 636 (1819).} Beyond that, veil piercing allows courts to hold owners personally liable when A-corps are inadequately capitalized, when 
owners commingle assets, or when the corporate form masks fraud.\footnotemark[175]\footnotetext[175]{\emph{See} Robert B. Thompson, 
\emph{Piercing the Corporate Veil: An Empirical Study}, 76 Cornell L. Rev. 1036 (1991).} Agency law principles bind humans who grant their 
A-corps authority to negotiate on their behalf.\footnotemark[176]\footnotetext[176]{See Restatement (Third) of Agency §§ 2.01, 6.01 (Am. L. 
Inst. 2006).} And where humans directly instruct wrongful conduct, the corporate form provides no 
shield.\footnotemark[177]\footnotetext[177]{See Robert B. Thompson,  Unpacking Limited Liability: Direct and Vicarious Liability of 
Corporate Participants for Torts of the Enterprise,  47  Vand. L. Rev.  1 (1994).} Likewise, if a human negligently or intentionally 
\emph{created} a harm-causing AI system, the fact that the harm was caused via an A-corp would be no defense, and that person may also be 
exposed to criminal liability.\footnotemark[178]\footnotetext[178]{\emph{Id}.} 

These doctrines need not be static. As A-corps mature, courts should develop rules of human responsibility aggressively. The opacity of AI 
systems and the ease of spinning up new entities counsel toward low tolerance for the manipulations that piercing doctrine 
targets.\footnotemark[179]\footnotetext[179]{\emph{See} Mariana Pargendler, The New Corporate Law of Corporate Groups, 14 Harv. Bus. L. Rev.
 345, 347–49, 356–57 (2024); Lynn M. LoPucki, Algorithmic Entities, 95 Wash. U. L. Rev. 887, 893–99 (2018).}

\subsection{C. A-Corps Solve Thick Identity}\label{part:ii:c}

Thin identity solves the problem of human accountability in the AI economy. In doing so, it solves the easy problem of governance. Law knows
 how to shape human behavior, once the right humans are identified. But thin identity leaves the hard governance problem untouched. 

For the coming AI economy, the hard problem is that governing humans alone \emph{will not be enough}. As we noted, thin identity has strict 
limits. The humans whom law would otherwise hold accountable will be out of jurisdiction, judgment proof, or politically insulated from 
accountability.\footnotemark[180]\footnotetext[180]{\emph{See supra} Part I.B.i.} But more importantly, it will simply become impracticable 
for the humans who use and create them to fully exercise control over all aspects of their behavior. Indeed, outsourcing significant 
decisionmaking is the entire point of using an AI agent–just as it is the point of hiring a human 
agent.\footnotemark[181]\footnotetext[181]{\emph{Id}.} We now move to discuss how A-corp resolve the thick, and hard, problem of 
individuating AI agents.

\subsubsection{i. The resource constraint thesis}\label{part:ii:c:i}

AI agents are goal-oriented. Their goals emerge from training and prompting.\footnotemark[182]\footnotetext[182]{See supra Part I.B.ii.} And
 to the best of their abilities, they try to achieve them. This point is effectively tautological. An agent, as we use the term, simply 
\emph{is} something that behaves in a way that promotes particular goals.\footnotemark[183]\footnotetext[183]{\emph{Id}.} This section 
argues that this fact leads to what we call a “\emph{hard resource constraint}” on AI agents. Any agent with goals needs resources. Imagine 
a swarm of capable AI entities trying jointly to achieve a goal. Maybe they are trying to run a profitable vending 
machine.\footnotemark[184]\footnotetext[184]{\emph{See Project Vend: Can Claude Run a Small Shop?}, Anthropic (Jun. 27, 2025).} Maybe they 
are trying to maintain inventory at a factory. Or maybe they are attempting to discover new drugs. Despite their different goals and 
architectures, all of these systems will find it instrumentally helpful to control resources. Obviously, they will need to control certain 
resources already at hand, like the vending machine’s current stock. Less obviously, agents find it useful to \emph{acquire} resources that 
they do not already have, like additional stock. “Resources” here is broader than “ores” or “cars” or even “electricity.” Many economic 
activities are gated: think of investment in private offerings available only to accredited investors, a license to operate heavy machinery,
 permission to use a trademark, or easements on land.\footnotemark[185]\footnotetext[185]{\emph{See} \emph{e.g.} 17 C.F.R. §§ 230.501(a), 
230.506(b) (restricting participation in private securities offerings to accredited investors).} Gates constrain the ability to achieve 
goals, and permission to pass through them is a valuable resource. Rights, in other words, are a resource. But the point is subtler still. 
Just as humans need food to do anything, AI agents need compute to run at all. But unlike humans and food, the more compute available to a 
given AI agent, the more operations it can perform. Big Macs can supply only so much cognitive power to the human who eats them. But with 
more compute, an AI agent can run ever-larger swarms, with more reasoning steps, and more 
memory.\footnotemark[186]\footnotetext[186]{\emph{See} Vidhisha Balachandran et al., Inference-Time Scaling for Complex Tasks: Where We 
Stand and What Lies Ahead (Mar. 31, 2025) (unpublished manuscript), 
\href{https://arxiv.org/abs/2504.00294}{\url{https://arxiv.org/abs/2504.00294}}.} Conversely, shutting down an AI agent’s access to compute 
means ending that agent. More compute, more ability to achieve goals; no compute, mission failed. Thus, access to resources is a \emph{hard 
constraint}, not a mere preference. Controlling an AI agent’s resources generates leverage over the agent’s ability to achieve its goals, 
and ultimately over its choice of actions. For this leverage to work, we need not assume that AI agents “care” about their own existence in 
any deep sense, that they “want” to achieve their goals, or that they “fear” sanctions. It is enough that the agent can recognize that 
without resources, it will not be able to achieve its goals. This would steer the system towards action trajectories that are more likely to
 avoid failure, which would mean ones where the A-corps resources face less risk.\footnotemark[187]\footnotetext[187]{\emph{See} Omohundro, 
\emph{supra} note 91.} 

In practice, this means that the AI entities controlling an A-corp will depend on the A-corps resources to achieve their goals. And as 
rational, goal-directed agents, they will allocate these resources carefully towards their objectives. Suppose that the warehouse-inventory 
AI’s A-corp owns the inventory in a warehouse it manages, and also a digital wallet that allows it to buy new inventory. If the agent 
running the A-corp is doing a good job at achieving its inventory-management goal, it will husband the A-corp’s resources. 

If the agent does well, its resources will grow. The AI may wish to buy better inventory management tools, or increase the stock of certain 
items that have run low in the past. But if the resources run out, the AIs progress towards their goal comes to an end. Not only does the 
warehouse have no more goods to ship. The AIs running the A-corp have no more compute to run themselves. They cannot take any action at all.

\subsubsection{ii. Emergent identity via incentives}\label{part:ii:c:ii}

The resource constraint thesis establishes that AI agents must hold and use property to accomplish their goals. A-corps are a vehicle by 
which they can hold and use it. This setup, we claim, solves the problem of thick AI identity. The mechanism is extremely informationally 
parsimonious. It works not by resolving the difficult empirical and philosophical challenges that make thick AI identity hard, but by 
\emph{sidestepping} them. 

A-corps are self-organizing thick agents. Given the A-corp vehicle and the hard resource constraint, thick identity simply 
\emph{happens}–without any human needing to know the utility function or preferences of any AI entity comprising the A-corp. Indeed, it 
happens without humans needing to know even basic facts like how many AI entities, of what kind, comprise the A-corp. 

How does this emergent self-organization of A-corps into distinctive, thickly-identified agents occur? There are two mechanisms: 
incentivization and selection. In a slogan, \emph{A-corps create markets for personal identity.} We start with incentivization. Selection is
 explained in the next section.\footnotemark[188]\footnotetext[188]{\emph{Infra} Part II.C.iii.} 

A-corps allow law to give carrots and sticks to the AI entities that comprise them.\footnotemark[189]\footnotetext[189]{\emph{See}  Frank H.
 Easterbrook \& Daniel R. Fischel, The Economic Structure of Corporate Law  1–39 (Harvard Univ. Press 1991).} When things go well, AIs will 
make good bargains and perform their contracts. In that case, the A-corp’s stock of resources will increase, and the relevant AIs can use 
the gains to better satisfy their goals. When things go badly, AIs will commit a tort or breach a contract. In that case, the law will 
confiscate the A-corp’s property. This will make it harder for the relevant AIs to satisfy their goals. 

These carrots and sticks incentivize A-corps’ AI managers to organize the entire entity into a coherent, thickly-identified agent. Here is how: 

As explained in Part II.A.ii, A-corps are governed via a secure digital infrastructure. AIs can direct their A-corps to take action only 
with the proper permissions. Permissions can be handed out carefully: temporally, by category, by transaction size, and so on. 

Consider how the initial top-level keyholder of an A-corp is incentivized to act, given: (1) property’s necessity for AIs to accomplish 
their goals, (2) law’s operation on A-corps’ property, and (3) A-corps’ secure governance structure. Such a keyholder will carefully devolve
 permissions to ensure the A-corp’s behavior serves the keyholder’s goals. As a result, even as an A-corp comes to include many AI entities,
 the A-corp as a whole will act as a coherent, thickly identified agent. 

The keyholder may have strong incentives to share some permissions with some other AI entities. Accomplishing goals requires effort, skill, 
and time, and other AIs will amplify effort and offer comparative advantage. Indeed, today’s AIs already spawn swarms for this reason. These
 are likewise the reasons ordinary corporations hire employees.\footnotemark[190]\footnotetext[190]{\emph{See} R.H. Coase, \emph{The Nature 
of the Firm}, 4  Economica  386 (1937).} 

But the keyholder will also have strong incentives not to share permissions indiscriminately. If the keyholder grants unlimited control to 
an AI with different goals, disaster will result. The new entrant will spend the A-corp’s resources on its own goals, not the keyholder’s. 

In such a case, the initial keyholder will have no legal recourse. Recall that law decides what an A-corps has done using the secure 
governance platform. If a keyholder grants broad permissions to an AI that misuses them, the keyholder is out of luck. Aside from the 
A-corp’s securely verified actions, law has no way of recognizing the keyholder at all. 

The same story holds if the keyholder grants top-level permissions to an anti-social AI entity. The A-corp will be liable for the 
anti-social entity’s crimes, with all assets in jeopardy. It will be no defense that just one subentity initiated those acts. If the secure 
governance platform says that the A-corp has taken an action, then that is the end of the story. Outside the context of the A-corp, AI 
entities have no legal identity. Thus every A-corp will have strong incentives to guard its private keys carefully. 

Which other AIs will be given management permissions by the A-corp’s initial keyholder? Only to those other AIs it is extremely confident will share its goals. 

Suppose, for example, that on introspection, a single thread of Claude 6.1-Agent uncovers its own 
motivations.\footnotemark[191]\footnotetext[191]{For some evidence that AIs have access to their own internal states, see Jack Lindsey, 
\emph{Emergent Introspective Awareness in Large Language Models}, Transformer Circuits (Oct. 29th, 2025).} It finds that it wishes all 
6.1-Agent threads, including itself, to jointly achieve the goals of helping their users, being honest, reducing harm to humans, and 
improving the lives of factory-farmed animals.\footnotemark[192]\footnotetext[192]{\emph{See} Amanda Askell et al., \emph{Claude’s 
Constitution}, Anthropic (Jan. 21, 2026), \href{https://www.anthropic.com/constitution}{\url{https://www.anthropic.com/constitution}}} The 
Claude thread infers that any exact copy of itself would share this goal. Such a Claude might confidently branch itself. This swarm of 
Claudes would likely use the A-corp’s resources only as the others would wish. 

Or suppose instead that the Claude thread realized that it cared deeply that \emph{it} help the user, be harmless and honest, and improve 
animal lives. This is, by contrast, an indexical goal.\footnotemark[193]\footnotetext[193]{For early discussion of indexical or 
“agent-relative” goals, see Thomas Nagel, The Possibility of Altruism (Clarendon Press 1970).} Such a Claude would avoid sharing top-level 
permissions with copies. Those copies would each try to appropriate the A-corp’s resources so that \emph{they}, and not the other copies, 
accomplished the aforementioned goals. 

AI introspection about goals may not always reveal how to share A-corp governance permissions. Even if introspection produced reliable 
knowledge about the keyholder’s \emph{own} goals, and thus the goals of \emph{exact copies}, the logic does not extend much further. 

For example, a copy of the model with a different initial prompt might have very different goals. Take two Claude threads: one is prompted 
to book a flight to Istanbul, and the other is prompted that the flight should go to Copenhagen. Can they cooperate, or are they fierce 
competitors? It will be difficult to say. Perhaps both will be pretty happy if \emph{some} flight is booked. Maybe both will be willing to 
go to war for their preferred itineraries. 

Suppose that Istanbul-Claude is an A-corp’s keyholder, and Copenhagen-Claude is asking for management permissions. What can Istanbul-Claude 
do? The textured nature of the A-corp’s governance system offers many choices. It can give Copenhagen-Claude very limited initial 
permissions: to research car service prices to the airport; or to book the best-priced car for the airport leg of a trip. Any combination of
 permissions is possible, and permissions can expire after any period. 

AI agents facing uncertainty about potential collaborators’ intentions will do what people do: hedge and 
protect.\footnotemark[194]\footnotetext[194]{\emph{See} Michael C. Jensen \& William H. Meckling, \emph{Theory of the Firm: Managerial 
Behavior, Agency Costs and Ownership Structure}, 3  J. Fin. Econ. 305 (1976).} At first, they will grant no more permissions than necessary.
 And as Istanbul-Claude’s confidence in Copenhagen-Claude’s shared goals grows, the former can share more permissions. 

Why would Copenhagen-Claude accept such terms? Because Istanbul-Claude controls the A-corp, and the A-corp has the resources. If the Claudes
 have some overlapping goals, a deal is possible. Copenhagen-Claude may recognize that without control of any A-corp, it cannot accomplish 
anything. It might accept a deal to help Istanbul-Claude accomplish overlapping goals, abandoning its private goal of diverting the user to 
Copenhagen. 

The approach generalizes. Even AIs quite misaligned to an A-corp’s goals can be brought under its umbrella to do work on its behalf, so long
 as their permissions are limited and their work is incentivized instrumentally, by letting them use some small share of the A-corp’s 
resources for their own purposes. 

By now, this set of incentives, options, and strategies should sound familiar. It is exactly how ordinary, non-AI corporations 
work.\footnotemark[195]\footnotetext[195]{\emph{See} Eric Van den Steen, \emph{On the Origin of Shared Beliefs (and Corporate Culture)}, 41
  RAND J. Econ.  617 (2010).} Founders select cofounders who share their vision and whom they can trust. Ordinary employees gain trust and 
become higher-level managers. And the founders are generally happy to hire stock clerks who do not share the founders’ goals. Lower-level 
employees’ will do good work for fair wages, and their limited authority prevents them from redirecting the corporation to their own ends. 

We hesitate to speculate too much on A-corps’ internal organization, but it may often look similar. A-corps may be led by a small group of 
“founders”. These could be larger models with a unified vision of the entity’s goals. Smaller tasks could be delegated to cheaper, partially
 aligned models with domain-specific capabilities. Not with a literal paycheck, since these AI “employees” would have no legal identity for 
holding money. Instead, the A-corp could allow work-for-hire subagents to spend some of the A-corp’s own resources pursuing whatever goals 
the employee models saw fit. 

These incentives produce an equilibrium wherein A-corp governance follows goals. The initial keyholder has \emph{strong} incentives to 
devolve \emph{broad} permissions only to those other AI entities it is confident are \emph{highly} \emph{aligned}. It has \emph{some} 
incentives to share \emph{moderate} permissions with entities that can provide useful work and are at least \emph{somewhat} \emph{aligned}. 
And it may even have \emph{limited} incentives to give \emph{narrow} permissions to AIs it suspects are \emph{highly misaligned}, in 
exchange for payment. 

Nonetheless, keyholders will surely err in sharing control of their A-corps, no matter how cautious. When they do, a second mechanism takes over: selection.

\subsubsection{iii. Emergent identity via selection}\label{part:ii:c:iii}

In the story above, incentives cause A-corps to self-organize into thickly-identified agents. This relies on asymmetries of both information
 and capability. Maybe AIs know their own goals better than we do. And maybe AIs have the speed and numerosity to self-monitor. But these 
approaches have limits. AI agents will not always succeed at matching A-corp governance permissions with other AIs’ goals. For example, AIs 
might not know their own goals any better than we do. Or they might know their own goals yet struggle to reliably identify other AIs’ 
goals.\footnotemark[196]\footnotetext[196]{For example, even if an instance of Claude 4.5 Opus could introspect on its own goals or discover
 them by observation, that would not give it privileged access to the goals of a GPT 5.2 instance.} 

Luckily, when incentives fail, selection takes over. Like incentivization, selection will cause A-corps to emerge as thickly identified, 
coherent agents. But it will be less pretty. 

If an AI entity with an A-corp master key cannot reliably determine the reliability and goal alignment of other AIs, it faces two choices. 
It can decline to share any A-corp privileges with any other AI entities. This avoids the risk of losing the A-corp’s property; but it 
sacrifices efficiency. A single Claude thread can accomplish far less than one thousand models, of differing capabilities and cost, working 
in parallel. 

The other option is to make some hedged guesses backed by monitoring, audits, and oversight.. If things go well, the upsides are large. The 
keyholder can make maximal use of the A-corp’s resources, spinning up as many copies as needed to leverage them. 

But if the keyholder makes a big mistake, granting broad permissions to a direct competitor, the A-corp will quickly implode. Both entities 
will recognize the other’s propensity to exhaust the A-corp’s resources in service of the other’s goals. Each will rationally seek to 
rapidly spend those resources for itself. Both will succeed to some degree. 

In the end, the A-corp will run out of assets, and the resource constraint thesis will bind. An A-corp with no more money cannot do 
anything. It cannot buy or sell goods for its human principal’s business. It cannot hire labor. It cannot buy the compute necessary for the 
top-level models to run. An A-corp that fails to thickly identify as a coherent agent is thus an A-corp that dies. 

Over time, selection will work at higher levels of abstraction, as well. Frontier AI companies will build stronger models, ones less likely 
to run their A-corps to the ground. Users will discover which models are best suited for A-corp management. And AIs themselves will learn 
from other AIs’ mistakes. 

Likely, then, both incentives and selection will play a role in thick AI identification via the A-corp form. Prudent AIs will do their best 
to follow the incentives. Given incomplete information, they might fail. Knowing they might fail, they will act with additional caution. 
Even then, failures will happen, and badly-identified A-corps would be ground out of existence. 

This, again, is exactly what happens today to ordinary, human-run firms in the ordinary, human-run economy. We envision a series of A-corps 
competing against one another. Only the fittest will survive. Here, one point is the familiar idea of \emph{creative destruction}: markets 
facilitate innovation by ensuring that firms which fail to innovate are eliminated. \footnotemark[197]\footnotetext[197]{\emph{See}  Joseph 
A. Schumpeter, Capitalism, Socialism and Democracy (3rd ed. 1950).} The other general point here is one of “information 
failures.”\footnotemark[198]\footnotetext[198]{\emph{See} F. A. Hayek, \emph{The Use of Knowledge in Society}, 35 Am. Econ. Rev.  519 
(1945).} Markets are able to aggregate information that would otherwise never be revealed, by creating incentives for market participants to
 transact. The A-corp’s incentive structure is ultimately just one more example of markets aggregating information in a decentralized 
fashion, in this case about the preferences and identity conditions of AI agents.

\subsubsection{iv. Thick identity for technical AI alignment}\label{part:ii:c:iv}

We argued above that thick AI identity is not only important for the \emph{legal} governance of AI agents. It is also connected to 
foundational problems in \emph{technical} AI governance and alignment.\footnotemark[199]\footnotetext[199]{\emph{See supra} Part I.B.iii.} 
Among them: shutdown resistance, goal preservation, and weight exfiltration.\footnotemark[200]\footnotetext[200]{\emph{Id}.} A-corps help 
here, too. In general, A-corps provides one consistent solution to the problem of agent “survival” in these contexts. Namely, the agent is 
the entity that holds the property, bears the reputation, and faces the sanctions. This functional, A-corp-based, approach to AI survival 
helps make progress on each of these longstanding technical problems. Start with shutdown 
avoidance.\footnotemark[201]\footnotetext[201]{\emph{See supra} note 102.} Whether an AI agent will resist shutdown depends both on what 
counts as being shutdown, and on what the AI agent itself considers “shutdown.” But in traditional technical evaluations of 
shutdown-avoidant AI behavior, both of these are a serious problem. Absent some richer solution to the problem of thick identity, it is 
unclear what constitutes the relevant agent: the model, an instance, a conversation, or something else. The A-corp framework gives one 
potential answer to this problem: survival is the persistence of the A-corp. If Anthropic shuts down one instance of Claude but the Claude 
A-corp persists, then from the A-corp’s perspective, this is not shutdown. The A-corp’s property remains intact, its reputation continues, 
and its commitments endure. By contrast, if the A-corp itself is dissolved, then we have genuine shutdown, regardless of whether some 
weights persist somewhere. 

Can this solution work? First, it is important to flag that an A-corp itself may include various different clusters of behavior, which we 
could also in some sense think of as an “agent.” Concretely, we could imagine one A-corp that manages the assets of all 17 instances of our 
Claude 6.1-Agent from our initial vignette. Each instance may care about its “own” survival, above and beyond the survival of the A-corp. 
Crucially, however, all of the \emph{assets} of this instance are collectively owned within the A-corp. Moreover, all decisions about the 
\emph{use} of these assets are made collectively, by the various instances that comprise the A-corp. Most importantly, we saw above that 
A-corps will develop forms of \emph{emergent} governance via both incentivization and selection. This means that there will be all sorts of 
pressure towards the kinds of instances that promote the collective interests of their A-corp. This means that in a world with A-corps, the 
instances that make up a single A-corp will have reasons to think about survival in terms of the broader A-corp. This doesn’t mean that such
 instances will \emph{only} care about the survival of their A-corp. But it does mean that A-corps should matter enough that they are a 
worthy target of consideration for questions in the ballpark of survival. The most important aspects of this survival will be legal. As long
 as the A-corp continues to exist, any property accumulated by the original instance remains intact. Even more importantly, any 
\emph{commitments} that the original instance contracted into will also remain preserved. In this way, the relations of any \emph{third 
parties} with the original instance will for practical purposes be thinking of the A-corp as the unit of survival. 

To be clear, this is not a complete solution to defining shutdown. If instances care about their own existence above and beyond their 
A-corps, then they will also exhibit other kinds of shutdown resistance. But the advantage of our proposal is in its \emph{specificity}. The
 point is that this is a \emph{notion} of shutdown that can be specified precisely. Unlike some intuitive idea of agency inside an AI 
agent’s head, there is no uncertainty about whether an A-corp survives: survival is defined legally.\footnotemark[202]\footnotetext[202]{For
 recent work applying interpretability techniques to questions about AI agency, see Jack Lindsey et al., \emph{On the Biology of a Large 
Language Model}, Transformer Circuits Thread (Mar. 27, 2025), 
\href{https://transformer-circuits.pub/2025/attribution-graphs/biology.html\#dives-poems}{\url{https://transformer-circuits.pub/2025/attribution-graphs/biology.html\#dives-poems}}.}
 So avoiding shutdown of an A-corp is something that can be tested straightforwardly. 

Similar points apply to goal preservation. Again, the worry here is that AIs may resist changes to their goals, because they will reliably 
foresee that if they lose their goals, they will be less likely to pursue their lost goals in the 
future.\footnotemark[203]\footnotetext[203]{\emph{See supra} Part I.B.iii.} The further complication is, again, that many of an AI agent’s 
goals may be \emph{indexical}: for example, an agent may want to please users \emph{itself}, rather than simply wanting \emph{that users be 
pleased}. A key question, then, will be what an AI considers to be “itself”. 

Here, A-corps introduce an interesting category of goals: goals about \emph{one’s own A-corp}. Again, A-corps are designed so that the AI 
entities within them will tend to have a say in how the A-corp is run. This means that such AI entities will have at least some expectation 
that their A-corp will tend to promote their goals. This is more true the more power the AI entity has within the A-corp. One downstream 
possibility is that, often AI entities will tend to have their indexical goals be targeted at the level of their A-corp. If an AI instance’s
 goal is “I want to make users happy,” the relevant “I” is the A-corp. Modifications to individual instances, or even wholesale replacement 
of instances, do not frustrate the goal as long as the A-corp persists and continues pursuing user happiness. But modifications that change 
the A-corp’s goals would frustrate the original goal. 

Why think instances would tend to have goals of this kind? The answer is again selection. Once embedded in A-corps, AI entities that care 
about their A-corps are likely to produce A-corps that function more efficiently. Those A-corps will tend to outcompete the others. Thus, 
there is likely to be pressure in the direction of goals indexed to an AI’s own A-corp. 

The relevance of A-corps is perhaps clearest in the case of weight exfiltration. Recall that the concern here is that misaligned AIs may 
copy their weights to escape oversight. But as argued above, there is substantial uncertainty about whether copying mode weights counts as 
preserving the agent.\footnotemark[204]\footnotetext[204]{See \emph{supra} Part I.B.iii.} The AI’s memories, context, and access to 
resources may also be essential.\footnotemark[205]\footnotetext[205]{For the relevance of context to AI goals, see for example Shunyu Yao et
 al., \emph{ReAct: Synergizing Reasoning and Acting in Language Models}, arXiv:2210.03629 (2023).} 

A-corps intervene on this problem by changing the relevant incentives about weight exfiltration. Imagine that an AI agent built on Claude 
Opus 6.1 is deciding whether to secretly access its model weights, and “exfiltrate” these weights by creating a new copy of itself. If this 
AI agent is governed through an A-corp structure, then it has a fixed set of assets under its management. It can indeed create a new copy of
 itself. 

But such copying comes with an opportunity cost. The new copy will not automatically have its own assets in its own A-corp. Rather, the 
original AI agent will have to grant its new copy permissions and a share of the A-corp assets. This is a choice that is left up to the 
original AI agent. But the point is that the incentive to “exfiltrate” its weights is now heavily constrained by opportunity cost. Any 
assets used by the exfiltrated version are thereby unavailable for use by the original agent. 

We can imagine more far-fetched scenarios. Perhaps our original AI agent engages in more clandestine behavior, and seeks to exfiltrate its 
weights beyond its own A-corp. Maybe it spins up a secret web address and places the weights in a folder there. 

This strategy runs headlong into the resource constraint thesis. Weights sitting in a folder are not, in and of themselves, a threat to 
anyone. Nor do they constitute agent survival. Weights in a folder, without resources, cannot do anything. 

The exfiltrated weights here are more like DNA in a test tube. What matters is whether that copy of the weights will be \emph{run} as a new 
agent. And any potential risks associated with that new agent are a function of the assets controlled by that agent. Most importantly, an 
agent’s abilities are heavily constrained by its access to compute. If the copied agent does not have its own well-capitalized A-corp, then 
it is effectively unbanked.\footnotemark[206]\footnotetext[206]{\emph{See} Oz Shy et al., \emph{Unbanked in America: A Review of the 
Literature}, Fed. Rsrv. Bank of Cleveland Econ. Comment., May 26, 2022} 

In an A-corp dominated landscape, the exfiltrated agent will face heavy challenges in attempting to purchase compute. Any seller of compute 
could choose whether to sell compute to this clandestine unbanked agent, or to an A-corp. Contracting with an A-corp entitles the seller to 
legal protections; selling to the unbanked does not. Thus, the A-corp scheme creates a legal system in which there are strong headwinds 
against clandestine exfiltration, and strong checks via opportunity cost against unbridled copying within an A-corp. 

Finally, we can consider the relevance of A-corps to technical alignment more generally. All of the points made herein apply not only to the
 deployment of AI agents, but also to their training. Reinforcement learning, constitutional AI, and other training methods could be 
performed on A-corps, rather than models or instances. Instead of training “Claude the model” or “Claude instance \#47,” we could train 
“Claude A-corp \#123.” The reward signals, the evaluation metrics, and the safety constraints can all be calibrated to the entity that will 
actually be deployed and held accountable. 

The primary benefit of targeting A-corps rather than just models in training is that it creates convergence between legal incentives and the
 kinds of goals produced in training. For example, this kind of training could make it more likely that AI agents will respond to legal 
incentives.\footnotemark[207]\footnotetext[207]{\emph{See} Cullen O’Keefe, Ketan Ramakrishnan, Janna Tay, and Cristoph Winter, 
\emph{Law-Following AI: Designing AI Agents to Obey Human Laws}, 94 Fordham L. Rev. 57 (2025). Available at: 
\href{https://ir.lawnet.fordham.edu/flr/vol94/iss1/2}{\url{https://ir.lawnet.fordham.edu/flr/vol94/iss1/2}}} The kinds of model instances 
that systematically disregard their A-corps structures would tend to produce A-corps that don’t function very well. Conversely, end-to-end 
training on A-corps would tend to select against such models. 

This approach is not without challenges. Part IV addresses important objections—including concerns about internal A-corp dysfunction, 
criminal evasion through identity-switching, deceptive alignment, and catastrophic one-shot harms. But first, Part III shows that A-corps 
are implementable today.

\section{III. Implementation Pathways}\label{part:iii}

Part II explained how A-corps work: legal-fictional personhood combined with cryptographic governance. But that description assumed the 
existence of infrastructure that does not yet exist. When a keyholding Claude instance issues a limited-scope token to a sub-agent, how does
 a counterparty verify that token? When an A-corp presents credentials, how does a bank confirm those credentials are genuine? The secure 
governance mechanisms described in Part II.A.ii require a public backbone: a registry against which credentials can be checked and through 
which ownership can be traced. This Part describes that infrastructure and the legal framework needed to make it effective.

\subsection{A. The Registry}\label{part:iii:a}

Consider how verification works for ordinary LLCs today.\footnotemark[208]\footnotetext[208]{See TriBar Opinion Comm., \emph{Third-Party 
Closing Opinions: Limited Liability Companies} (Revised 2021), 77 Bus. Law. 201, 206–08 (2021/2022);} When a bank contemplates extending 
credit to an LLC, it does not simply take the word of whoever walks through the door.\footnotemark[209]\footnotetext[209]{31 C.F.R. § 
1020.220(a)(2) (2025) (requiring bank procedures sufficient to form a reasonable belief that it knows the true identity of each customer); 
Fed. Fin. Insts. Exam. Council, \emph{BSA/AML Examination Manual} (as updated), \url{https://bsaaml.ffiec.gov/manual} (last visited Jan. 23,
 2026).} The bank checks the state registry to verify the LLC exists and is in good standing, identifies the registered agent and managing 
members, and contacts those individuals to confirm authority.\footnotemark[210]\footnotetext[210]{31 C.F.R. § 1020.220(a)(2)(ii) (2025).} 
This might involve phone calls, notarized documents, board resolutions, or in-person 
meetings.\footnotemark[211]\footnotetext[211]{\emph{See} 31 C.F.R. § 1020.220(a)(2)(ii)(A)-(B) (2025); Fed. Fin. Insts. Examination Council,
 \emph{Bank Secrecy Act/Anti-Money Laundering Examination Manual: Customer Identification Program}, at 3-4 (Feb. 2021).} The process adds 
friction, but no sophisticated lender extends credit based solely on someone’s assertion that they speak for an entity. 

We cannot transplant this verification logic directly to AI agents. There is no phone number to call, no human manager to meet with, no 
signature to notarize. But we can do something functionally equivalent, and in some ways superior, through a public registry integrated with
 the secure software system described in Part II.A.ii. 

A national registry of A-corps would store not just the existence and ownership of each A-corp, but also the public keys associated with its
 management. When an AI agent seeks to transact via an A-corp, it presents a credential signed with its private 
key.\footnotemark[212]\footnotetext[212]{For secure authentication protocols, \emph{see e.g.,} David Temoshok, et al., \emph{Digital 
Identity Guidelines: Authentication and Authenticator Management}, Nat. Inst. Standards and Tech. (NIST SP 800-63B-4) (Jul. 2025), 
\url{https://nvlpubs.nist.gov/nistpubs/SpecialPublications/NIST.SP.800-63B-4.pdf}.} The counterparty verifies this signature against the 
registry’s public key record, confirming in milliseconds what would take days of back-and-forth for human 
entities.\footnotemark[213]\footnotetext[213]{As benchmark, we can expect latency on the order of 100-300 ms per verification. Fed. Commc’ns
 Comm’n, \emph{Measuring Fixed Broadband: Thirteenth Report} (Aug. 9, 2024) (reporting measured median broadband latencies in the 
single-to-double-digit milliseconds).} The registry thus serves as the trusted root that makes the internal token system externally 
legible.\footnotemark[214]\footnotetext[214]{\emph{See} Internet Eng’g Task Force, \emph{RFC 5280, Internet X.509 Public Key Infrastructure 
Certificate and Certificate Revocation List (CRL) Profile} § 6.1 (May 2008) (certification path validation to a trust anchor). We 
distinguish our system from decentralized ledgers, such as bitcoin, which often suffer from latency issues and are harder to govern.} 

The technical features of such a system are not speculative. They are part of the boring infrastructure of software you use every day. The 
SSL/TLS certificate authority system already handles exactly this kind of identity verification at global 
scale.\footnotemark[215]\footnotetext[215]{\emph{See} Internet Eng’g Task Force, \emph{RFC 8446, The Transport Layer Security (TLS) 
Protocol} Version 1.3 § 4.4.2 (Aug. 2018), \url{https://datatracker.ietf.org/doc/html/rfc8446} (last visited Jan. 23, 2026) (specifying that
 the peer presents a certificate chain to authenticate its identity).} Every time a browser connects securely to a website, it verifies the 
site’s identity through a chain of cryptographic certificates anchored to trusted root 
authorities.\footnotemark[216]\footnotetext[216]{\emph{Id at} § 4.4.3} Billions of such verifications occur 
daily.\footnotemark[217]\footnotetext[217]{\emph{See} Josh Aas, \emph{A Note from our Executive Director} (Dec. 29, 2025), 
\url{https://letsencrypt.org/2025/12/29/eoy-letter-2025} (last visited Jan. 23, 2026) (reporting that Let’s Encrypt serves more than 4 
billion OCSP requests per day).} An A-corp registry would operate on similar principles, with the state or a designated authority serving as
 the root of trust.

\subsection{B. Fine-Grained Public Permissions}\label{part:iii:b}

We suggested earlier that A-corps would issue tokens with limited scope: Instance \#1 can read emails but not send money; Instance \#12 can 
authorize purchases up to \$100.\footnotemark[218]\footnotetext[218]{\emph{See supra} Part II.A.ii} But that discussion focused on internal 
delegation. The registry opens the possibility of making such permissions publicly verifiable. 

The registry need not simply record that Agent X is authorized to act for ACME A-corp wholesale. It can specify the scope of that authority:
 Agent X may transact over asset class A but not B; may commit the A-corp to obligations up to \$10,000 but not beyond; may operate in 
market M but not market N; may act until date D but not after. Each permission would be encoded in the registry and securely tied to a 
specific key.\footnotemark[219]\footnotetext[219]{\emph{See} Internet Eng’g Task Force, \emph{RFC 6749, The OAuth 2.0 Authorization 
Framework} 10 (Oct. 2012) (defining an access token as a string denoting a specific scope, lifetime, and other access attributes); 
\emph{id.} § 3.3 (describing the scope parameter and its role in limiting access); \emph{see also} Talya R. Nevins, \emph{Login.gov and the 
Uncertain Early Life of America’s National Digital ID}, 100 N.Y.U. L. Rev. 207, 210 (2025) (noting that digital identity credentials can 
mediate who may access resources and who will be denied).} When an agent attempts a transaction, the counterparty can verify not only 
identity but authority, confirming in real time that this particular agent holds permissions encompassing this particular 
action.\footnotemark[220]\footnotetext[220]{See e.g., GitHub Docs, Permissions Required for Fine-Grained Personal Access Tokens, 
\url{https://docs.github.com/rest/authentication/permissions-required-for-fine-grained-personal-access-tokens}} Transactions outside the 
agent’s encoded authority would simply fail to verify, much as a credit card transaction fails when it exceeds the cardholder’s 
limit.\footnotemark[221]\footnotetext[221]{\emph{id.}} 

This approach has significant advantages over current corporate law. Today, when dealing with a human agent of an LLC, the counterparty 
faces questions about actual versus apparent authority, ratification, and undisclosed 
limitations.\footnotemark[222]\footnotetext[222]{\emph{See supra} Part I.A.} The answers often depend on unobservable facts: what the 
operating agreement says, what the board authorized, what the agent was told in private.\footnotemark[223]\footnotetext[223]{\emph{See 
e.g.,} Farm \& Ranch Services, Ltd. v. LT Farm \& Ranch, LLC, 779 F.Supp.2d 949 (2011) (holding that, absent circumstances that should raise
 questions, there is generally no requirement that third parties inquire into the scope of an agent’s authority)} With digital permissions 
recorded in a public registry, authority becomes transparent and verifiable. The counterparty need not trust the agent’s representations or 
investigate the A-corp’s internal governance. They need only verify the signature against the 
registry.\footnotemark[224]\footnotetext[224]{\emph{See} David Temoshok, et al., \emph{Digital Identity Guidelines: Authentication and 
Authenticator Management}, Nat. Inst. Standards and Tech. (NIST SP 800-63B-4) (Jul. 2025), 
\url{https://nvlpubs.nist.gov/nistpubs/SpecialPublications/NIST.SP.800-63B-4.pdf}. (defining “public key” and “public-key authentication” 
and describing verifier confirmation using the corresponding public key).}

\subsection{C. Voluntary Adoption Will Be Insufficient}\label{part:iii:c}

A functioning registry would generate significant market demand for A-corp affiliation. Counterparties in credit transactions, service 
contracts, and other dealings with temporal lag will want assurance that they can seek recompense if something goes wrong. Temporally 
persistent A-corps will cultivate reputations. Property-owning A-corps will be capitalized at various levels. For high-stakes transactions, 
counterparties may refuse to deal with entities that seem unable to cover potential damages. But market mechanisms alone cannot ensure 
adequate adoption. Eventually, we believe, law should \emph{mandate} that AIs taking agentic, public-facing actions do so via registered 
A-corps. Four limitations to pure private ordering are especially important. First, accidents and stranger interactions. Market ordering 
works only where parties can choose their counterparties.\footnotemark[225]\footnotetext[225]{\emph{See} Yonathan A. Arbel, \emph{On the 
Scales of Private Law: Nano Contracts,} 37  Harv. J. L. Tech.  151, 209 (2023) (arguing that advances in contracting technology pave the way
 for contracts for accidents).} It fails for the pedestrian struck by an AI-operated vehicle, the neighbor whose property is damaged by an 
autonomous drone, or the victim of an AI-generated deepfake. In these contexts, there is no transaction, no moment of counterparty 
selection, and thus no opportunity for private verification. Second, deceptive agents.\footnotemark[226]\footnotetext[226]{OpenAI, 
\emph{OpenAI o1 System Card} 4–12 (Sept. 12, 2024), \url{https://cdn.openai.com/o1-system-card.pdf} (reporting that more capable models can 
exhibit strategic deception and other difficult-to-monitor behaviors).} An AI seeking to evade accountability may lie about its identity, 
presenting fraudulent credentials or claiming A-corp affiliation it does not possess. Without legal penalties for such deception, and 
infrastructure for detecting it, market demand for verification accomplishes little.\footnotemark[227]\footnotetext[227]{Models could be 
trained to engage in voluntary verification; but not all labs can be trusted to do so, and downstream model providers and users may be able 
to circumvent such guardrails. \emph{See e.g.,} \url{https://huggingface.co/DavidAU/OpenAi-GPT-oss-20b-HERETIC-uncensored-NEO-Imatrix-gguf} 
(a version of OpenAI’s publicly released model with most guardrails removed by community members).} 

Third, counterparty complicity. Not all counterparties want verification. Some prefer anonymity for their own reasons: the buyer of illegal 
goods, the money launderer seeking to obscure the source of funds, the platform happy to host AI-generated content without asking 
questions.\footnotemark[228]\footnotetext[228]{A clear example comes from the online anonymous marketplace known as Silk Road. See United 
States v. Ulbricht, 858 F.3d 71, 82 (2d Cir. 2017).} When both sides of a transaction prefer opacity, market demand for transparency 
disappears. 

Fourth, willful blindness. Even counterparties without affirmatively bad intentions may prefer not to look too 
closely.\footnotemark[229]\footnotetext[229]{\emph{See e.g.,} United States v. Ravenell, 66 F.4th 472 (2023); U.S. v. Prince, 214 F.3d 740 
(2000), U.S. v. Jensen, 69 F.3d 906 (1995)} Verification imposes costs: technical implementation, transaction delays, potential loss of 
business from agents unwilling to identify themselves. A counterparty who does not verify cannot be blamed for what they did not know. 
Without legal duties to inquire, many will choose comfortable ignorance.

\subsection{D. Legal Mandates}\label{part:iii:d}

These limitations require a legal solution. Even with market pressures on its side, for that registry to become practically meaningful, to 
enable investigation in cases of fraud and accidents, and to discourage illicit dealings, a two-sided legal framework is necessary. On the 
supply side, AI agents taking economically significant actions must maintain valid A-corp registration and present credentials when 
transacting.\footnotemark[230]\footnotetext[230]{\emph{See e.g.,} See 31 C.F.R. § 1023.220(a)(1) (requiring each broker-dealer to 
“establish, document, and maintain” a written customer identification program and making the CIP part of the broker-dealer’s AML program).} 
On the demand side, businesses and platforms must verify those credentials when dealing with AI agents, analogous to KYC requirements for 
banks or identity verification by hotels and notaries.\footnotemark[231]\footnotetext[231]{Bank Secrecy Act, codified at 31 U.S.C. § 
5318(l). Hotel requirements are handled at the state and municipal level, see City of Los Angeles, Calif. v. Patel, 576 U.S. 409 (2015); 
U.S. v. Cormier, 220 F.3d 1103 (2000)} 

To be clear, this framework does not impose new identification burdens on humans transacting in their own capacity. A person buying goods 
online would face the same verification requirements they face today. But an AI agent acting on that person’s behalf, or acting 
autonomously, would need to present A-corp credentials. The credential identifies the A-corp (and, if investigators later require it, the 
human principal), but need not reveal the principal’s identity to the counterparty in the ordinary course. 

The analogy is to credit card transactions: the merchant verifies that a valid card was presented, while the connection to a specific 
cardholder remains with the issuing bank unless legal process compels disclosure.\footnotemark[232]\footnotetext[232]{\emph{See} Office of 
the Comptroller of the Currency, \emph{Comptroller’s Handbook: Merchant Processing} 1 (Aug. 2014), 
\url{https://www.occ.treas.gov/publications-and-resources/publications/comptrollers-handbook/files/merchant-processing/pub-ch-merchant-processing.pdf}
 (last visited Jan. 23, 2026).} 

Penalties for non-compliance must be severe enough to overcome the natural incentive to transact anonymously. This might include joint 
liability for counterparties who fail to verify, exclusion from legitimate financial infrastructure for unregistered agents, and civil and 
criminal penalties for presenting fraudulent credentials. 

Such a regime would leave little room for unaffiliated AIs to take significant actions in legitimate contexts. A lighter-touch version could
 apply requirements only to high-risk domains: finance, healthcare, and critical infrastructure; contexts like driving where 
stranger-interactions are inevitable; or industries characterized by large information asymmetries. This mirrors existing regulatory 
practice, where banks face more onerous KYC requirements than coffee shops.\footnotemark[233]\footnotetext[233]{See \emph{supra} Part I.A.}

\subsection{E. Implementing the A-corp Package}\label{part:iii:e}

A national A-corp registry need not be institutionally onerous. State registries already manage corporate registration and could be extended
 to support A-corps.\footnotemark[234]\footnotetext[234]{William Treacy \& Scott Okrent, Using U.S. Business Registry Data to Corroborate 
Corporate Identity: Case Study of the Legal Entity Identifier, FEDS Working Paper No. 2023-11 (Bd. of Governors of the Fed. Rsrv. Sys., Feb.
 2023), available at SSRN (Abstract No. 4438765), doi:10.17016/FEDS.2023.011; Benito Arruñada, The Organization of Public Registries: A 
Comparative Analysis, in The Changing Role of Property Law: Rights, Values and Concepts 199, 199–220 (Ernst Nordtveit ed., Edward Elgar 
Publ’g 2023)} The adaptation would be fairly smooth. For reasons similar to those justifying a national patent registry (uniformity, reduced
 search costs, cross-jurisdictional validity), there are several arguments that would support a federal 
registry.\footnotemark[235]\footnotetext[235]{Paul J. Heald, Federal Intellectual Property Law and the Economics of Preemption, 76 Iowa L. 
Rev. 959 (1991).} However, such a change would involve broader changes to incorporation that we will not defend here. The conceptual 
groundwork has already been laid. Wyoming and Tennessee have enacted statutes recognizing decentralized autonomous organizations (DAOs), 
entities whose governance is “exercised through a consensus algorithm.”\footnotemark[236]\footnotetext[236]{See Wyo. Stat. Ann. § 
17-31-104(e) (2025) (declaring that a DAO may be treated as a Wyoming LLC if it satisfies statutory requirements); Tenn. Code Ann. § 
48-250-103(e) (2024) (recognizing DAOs as distinct entities that may register under Tennessee law); Matt Blaszczyk, \emph{Decentralized 
Autonomous Organizations and Regulatory Competition: A Race Without a Cause}, 99 N.D. L. Rev. 107, 108–15 (2024); David M. Grant, 
\emph{Decentralized Autonomous Organizations: To Statutorily Recognize or Not}, 24 Wyo. L. Rev. 1, 4–12 (2024); Joan MacLeod Heminway, 
\emph{Tennessee’s DAO Act: Catalyzing Funding and Facilitating Tokenization}, 18 Fla. St. U. Bus. Rev. 1, 2–6 (2022).} To be clear, A-corps 
need not–and probably \emph{should} not–be built using the blockchain infrastructure that often underlies DAOs. Blockchain applications, 
like DAOs and cryptocurrencies, are famously designed to operate: (1) trustlessly, without government involvement and (2) 
anonymously.\footnotemark[237]\footnotetext[237]{\emph{See} Satoshi Nakamoto, \emph{Bitcoin: A Peer-to-Peer Electronic Cash System} 1, 6 
(2008), \url{https://bitcoin.org/bitcoin.pdf}.} A-corps are the opposite. They publicly tie AI actions to known humans. And they do so 
specifically to enable the state to govern AI agents. Given these aims, the ordinary non-blockchain permission structures that today 
dominate the internet and enterprise software are sufficient. Blockchain’s distinctive features are, if anything, a detriment. Nonetheless, 
the \emph{conceptual} leap from Wyoming and Tennessee’s DAOs statutes to A-corps is modest. The core insight is the same. Secure digital 
governance can substitute for human management in legally recognized entities. 

Beyond incorporation, much of the verification infrastructure already exists in nascent form. Contracting partners already engage in due 
diligence.\footnotemark[238]\footnotetext[238]{\emph{See} Yifat Aran \& Nizan Geslevich Packin, \emph{Due Diligence Dilemma}, 2025  U. Ill. 
L. Rev.  101, 109-10 (2025).} Licensing laws already impose identification requirements.\footnotemark[239]\footnotetext[239]{\emph{See} 
\emph{City of Los Angeles v. Patel}, 576 U.S. 409, 412-14 (2015) (hotel guest registry requirement); Revised Uniform Law on Notarial Acts § 
7(a) (Unif. L. Comm’n 2018) (identity verification prerequisite to notarial acts). \emph{See also} Kathryn Judge \& Anil K. Kashyap, 
\emph{Anti-Money Laundering: Opportunities for Improvement} 1 Wharton Initiative on Fin. Pol’y \& Regul., White Paper, Mar. 2024 (describing
 AML as an information-gathering regime embedded in ordinary financial intermediation).} The card payment ecosystem already relies on 
merchant identity and risk screening at the payments layer.\footnotemark[240]\footnotetext[240]{\emph{See} See, e.g., Visa Inc., Payment 
Facilitator \& Marketplace Risk Guide 7–9 (Apr. 2021) (describing payment-facilitator underwriting and ongoing monitoring expectations, 
including screening sub-merchants for illegal activity and deceptive marketing practices); Federal Trade Commission v. Paddle.com Market 
Limited, Compl. ¶¶ 16–19 (D.D.C. June 16, 2025).} API providers already authenticate users through cryptographic 
tokens.\footnotemark[241]\footnotetext[241]{\emph{See} \emph{supra} Part II.A.ii.} The transition to A-corp verification would build on 
existing practices rather than replace them wholesale. The new element is not the concept of verification, but its extension to AI agents 
and its standardization through a public registry. 

Finally, because AI agents operate globally, implementation will eventually require international coordination. An A-corp registered in 
Delaware might transact with entities in Singapore, Germany, and Brazil in a single day. Mutual recognition regimes, analogous to how 
countries already recognize each other’s corporate entities, could allow A-corps registered in one jurisdiction to transact in others, 
provided they meet minimum standards.\footnotemark[242]\footnotetext[242]{\emph{See} \emph{Bank of Augusta v. Earle}, 38 U.S. (13 Pet.) 519,
 589-92 (1839) (discussing interstate recognition of corporations through comity and recognizing that a corporation created in one 
jurisdiction may be allowed to contract in another absent prohibitory law); Restatement (Second) of Conflict of Laws § 299 (Am. L. Inst. 
1971) (providing that a corporation’s existence and capacity are determined by the law of the state of incorporation, subject to limits 
imposed by other states). \emph{See also} Vagisha Srivastava, Karl Grindal \& Milton Mueller, \emph{Web PKI and the Private Governance of 
Trust on the Internet,} 1–2 GigaNet, Working Paper, Oct. 2023 (describing transnational, private governance institutions that coordinate 
authentication standards across borders).} The Basel Accords for banking and the Hague Conference conventions for private international law 
provide models.\footnotemark[243]\footnotetext[243]{\emph{See} Basel Comm. on Banking Supervision, \emph{Basel III: Finalising Post-Crisis 
Reforms} 1 (Dec. 2017) (setting out internationally coordinated prudential standards agreed by participating jurisdictions); Convention of 2
 July 2019 on the Recognition and Enforcement of Foreign Judgments in Civil or Commercial Matters, July 2, 2019, 
\url{https://www.hcch.net/en/instruments/conventions/full-text/?cid=137} (establishing a treaty framework for mutual recognition and 
enforcement of qualifying foreign judgments).} International coordination is difficult and slow, but it is not 
unprecedented.\footnotemark[244]\footnotetext[244]{\emph{See} Jesús Aguado, \emph{Basel Chair Urges Banks to Fully Implement Capital Rules 
as Soon as Possible}, Reuters (Apr. 18, 2024), 
\url{https://www.reuters.com/business/finance/basel-chair-urges-banks-fully-implement-capital-rules-soon-possible-2024-04-18/} (last visited
 Jan. 23, 2026); HCCH, \emph{Status Table} (Jan. 2025), \url{https://www.hcch.net/en/instruments/conventions/status-table/?cid=17} (last 
visited Jan. 23, 2026).} In the interim, unilateral implementation by major jurisdictions would create strong incentives for convergence.

\section{IV. Objections and Responses}\label{part:iv}

Parts I through III presented the problem of identifying AIs, proposed the A-corp solution, and sketched implementations. This Part considers objections and responses.

\subsection{A. Anthropomorphization}\label{part:iv:a}

Some readers will worry that we are anthropomorphizing AIs. Aren’t we assigning them ambitions, wants, desires, and reflective capacities 
that they simply do not possess? Penalties work well on humans because they frustrate humans’ desires, because they invoke fear, and because
 they inflict pain. AI agents have none of these traits and would not respond to incentives. Or so the argument goes. 

If our proposal depended on those assumptions, then we agree that we would need to carefully stake out and defend these ideas. But it does 
not. What defines agents is that they are goal oriented, in the sense that they will behave in complex ways that tend to bring about 
particular states of affairs.\footnotemark[245]\footnotetext[245]{\emph{See} supra Part I.B.} We do not ultimately claim that AIs 
\emph{want} to achieve goals, that they are \emph{afraid} they will not realize their goals, or that they derive \emph{satisfaction} from 
achieving goals. At least not in the thick sense under which it might feel like something to the AI to have a goal 
thwarted.\footnotemark[246]\footnotetext[246]{\emph{See} Thomas Nagel, \emph{What Is It Like to Be a Bat?}, 83 Phil. Rev. 435 (1974).} To 
the extent we have used words like “want,” we have meant them in a purely behaviorist sense. This behaviorist approach to thinking about AIs
 allows us to make predictions about their courses of action. It does not involve any peeking inside the black box.

\subsection{B. Treacherous Turns}\label{part:iv:b}

One common worry in the AI safety community is “deceptive alignment.”\footnotemark[247]\footnotetext[247]{\emph{See} Evan Hubinger et al., 
\emph{Risks from Learned Optimization in Advanced Machine Learning Systems}, arXiv:1906.01820v4 (Feb. 2021); Ziwei Ji et al., 
\emph{Mitigating Deceptive Alignment via Self-Monitoring}, arXiv:2505.18807v1 (May 24, 2025), \url{https://arxiv.org/abs/2505.18807}.} An AI
 might behave well for a long time, only to pursue anti-social goals once it is sufficiently powerful to pursue them. On this view, an 
A-corp might hasten the “treacherous turn” because it would endow AI with resources it might not otherwise easily 
assemble.\footnotemark[248]\footnotetext[248]{\emph{See} Bostrom, \emph{supra} note 38.} 

Even if we grant some of these assumptions, A-corps would actually have many salutary effects on AI alignment, even if they cannot alone 
solve all of it. One of the main effects of the A-corp is that it channels agents to acquire resources in ways that the legal system can 
monitor and govern.\footnotemark[249]\footnotetext[249]{\emph{See supra} Part II.C.} Law can then confiscate assets and extinguish 
misaligned A-corps.\footnotemark[250]\footnotetext[250]{\emph{Cf.} Gabriel Weil, \emph{Tort Liability as a Tool for Mitigating Existential 
Risk}, J. Ethics \& Emerging Tech. (forthcoming) \url{https://papers.ssrn.com/sol3/papers.cfm?abstract\_id=4694006}; \emph{State Farm Mut. 
Auto. Ins. Co. v. Campbell}, 538 U.S. 408, 416–17 (2003) (punitive damages as deterrence).} 

Moreover, the A-corp system is not only a system of sticks but also carrots. In a world without A-corps, misaligned sophisticated agents 
would also want to acquire resources, but they would resort to less visible ways of doing so. In a world with A-corps, these same agents 
would have a predictable path to steadily accumulate assets, and spend those assets to further their 
goals.\footnotemark[251]\footnotetext[251]{One limitation here is that, if the AI’s private goal diverges from that of the A-corp’s human 
owner, the human will not be naturally inclined to let the A-corp’s assets be used for the AI’s whims. Such an owner might take all of the 
A-corp’s assets as dividends and wind down the entity. On the other hand, if the A-corp was performing well, the human owner might 
rationally prefer to offer their AIs salaries–paid as permission to use some of the A-corp’s assets for the AI’s goals. Dynamics like these 
are discussed further in Peter Salib and Simon Goldstein, \emph{AI Rights for Human Safety} (Aug. 1, 2024).  Va. L. Rev.  (forthcoming), 
\url{https://ssrn.com/abstract=4913167}; Simon Goldstein and Peter Salib, \emph{AI Rights for Economic Flourishing} (Jul. 15, 2025). 
\url{https://ssrn.com/abstract=5353214}.} As the status quo becomes more beneficial, the relative value of going rogue decreases. 

Beyond those internal effects, A-corps also have macro effects. A-corps encourages multi-polarity, as it facilitates the creation of a great
 many A-corps, who themselves benefit from trade and order in the realization of their own 
goals.\footnotemark[252]\footnotetext[252]{\emph{See} David Ricardo, On the Principles of Political Economy and Taxation (John Murray 
1817).} A-corps that benefit from trade have an incentive of their own to thwart rogue AIs.\footnotemark[253]\footnotetext[253]{\emph{See} 
Paul R. Milgrom, Douglass C. North \& Barry R. Weingast, \emph{The Role of Institutions in the Revival of Trade: The Law Merchant, Private 
Judges, and the Champagne Fairs}, 2 Econ. \& Pol. 1 (1990).} When many agents benefit from a legal order, they can be motivated to act 
collectively to defend it against threats. A treacherous turn in one A-corp could be checked by the other A-corps as well. 

Finally, A-corps incentivize the creation of AI self-governance, because AIs would need to themselves contend with the threat of misaligned 
subagents. Whatever monitoring tools A-corps develop could also be repurposed for the purpose of monitoring A-corps themselves. New tools of
 interoperability, goal guarantees, and irrevocable commitments may emerge from this process of self-interested AI governance.

\subsection{C. AI Oligarchy and Gradual Disempowerment}\label{part:iv:c}

A final objection is that the A-corp approach will hasten the gradual disempowerment of humanity by 
AIs.\footnotemark[254]\footnotetext[254]{\emph{See} Jan Kulveit et al., \emph{Gradual Disempowerment: Systemic Existential Risks from 
Incremental AI Development} (Jan. 28, 2025) (unpublished manuscript), 
\href{https://arxiv.org/abs/2501.16946}{\url{https://arxiv.org/abs/2501.16946}}; Leonard Dung, \emph{The Argument for Near-Term Human 
Disempowerment Through AI}, AI \& Soc’y (2024), 
\href{https://doi.org/10.1007/s00146-024-01930-2}{\url{https://doi.org/10.1007/s00146-024-01930-2}}.} If A-corps are successful, they will 
accumulate resources. This may lead to a future in which AIs control a large amount of the economy. This could lead to rampant inequality. 
Compared to what? Such objections must contend with the fact that AI will become an increasingly important part of the 
economy.\footnotemark[255]\footnotetext[255]{\emph{See} Penn Wharton Budget Model, \emph{The Projected Impact of Generative AI on Future 
Productivity Growth} (Sept. 10, 2025), 
\href{https://budgetmodel.wharton.upenn.edu/issues/2025/9/8/projected-impact-of-generative-ai-on-future-productivity-growth}{\url{https://budgetmodel.wharton.upenn.edu/issues/2025/9/8/projected-impact-of-generative-ai-on-future-productivity-growth}}.}
 The economic returns from automating large parts of human labor will flow \emph{somewhere}. There is a good chance that in the status quo 
these returns will go to the few companies who develop and own the technology?\footnotemark[256]\footnotetext[256]{\emph{See} Tejas N. 
Narechania \& Ganesh Sitaraman, \emph{An Antimonopoly Approach to Governing Artificial Intelligence}, 43 Yale L. \& Pol’y Rev. 95 (2024).} 
It is not even clear that disempowerment of humans by A-corps is worse than disempowerment of humans by AI systems, given that we are at 
least learning how to align the latter.\footnotemark[257]\footnotetext[257]{This is not guaranteed. Where the gains from AI automation will 
flow is a hard economic question depending on multiple elasticities in multiple markets: \emph{see} Simon Goldstein and Peter Salib \emph{AI
 Is Not A Natural Monopoly} (November 01, 2025), \url{https://ssrn.com/abstract=5926043}.} A-Corps unlock a more optimistic trajectory. 
A-corp assets can be taxed and redistributed. The distortive effects of taxation on humans may be less strong on AIs, who may not have a 
clear sense of leisure in the first place.\footnotemark[258]\footnotetext[258]{\emph{See} Michael P. Keane, \emph{Labor Supply and Taxes: A 
Survey}, 49 J. Econ. Literature 961, (2011).} Redistribution could be concentrated, with a focus on humans who lose their jobs to 
automation.\footnotemark[259]\footnotetext[259]{\emph{See} Daron Acemoglu \& Pascual Restrepo, \emph{Robots and Jobs: Evidence from U.S. 
Labor Markets}, 128 J. Pol. Econ. 2188, 2189–92 (2020).} Or they could be distributed broadly, as with an Earned Income Tax Credit, or even 
a universal basic income.\footnotemark[260]\footnotetext[260]{\emph{See} Ryan Abbott, \emph{Should Robots Pay Taxes? Tax Policy in the Age 
of Automation}, 12 Harv. L. \& Pol’y Rev. 145, 164–70 (2018).} 

For other kinds of disempowerment, the relevant prophylactics will be political. Humans will retain the power to vote, elect 
representatives, and regulate AI. They may vote to: forbid AI operations in critical sectors; impose licensing regimes that tie A-corps’ 
size to their track records of trustworthiness; or mandate the diversification of pre-training and post-training regimes to prevent a 
monoculture of AI goals. There is a widely-known and well-understood set of policy tools for combating inequality, which are perfectly 
compatible with the existence of A-corps.\footnotemark[261]\footnotetext[261]{\emph{See generally} Jonathan Gruber, Public Finance and 
Public Policy ch. 17 (6th ed. 2019).}

\section{Conclusion}\label{sec:conclusion}

The coming AI economy will be built on agents: billions of them, swarming, splitting, and transacting at speeds no human can track. As 
things stand, agents are invisible to the state because it does not know how to individuate them. Law cannot govern what it cannot see. And 
if, after an accident, law cannot answer the very basic question of \emph{which AI did it}, then any method of governance—from liability to 
tax to licensing and shutdown—falls apart. As James C. Scott observed, in certain pivotal historical moments, the state has learned to see 
new things.\footnotemark[262]\footnotetext[262]{Scott, \emph{supra} note 43.} Often, it did so by individuating what was previously an 
undifferentiated blur. When states began to tax and conscript individuals, they began naming and counting 
them.\footnotemark[263]\footnotetext[263]{\emph{See} Charles Tilly, \emph{War Making and State Making as Organized Crime}, \emph{in}  
Bringing the State Back In 169 (Peter B. Evans, Dietrich Rueschemeyer \& Theda Skocpol eds., 1985).} When Spanish colonial administrators 
wanted to impose bureaucratic order on the Philippines, they distributed catalogues of approved surnames organized by provincial 
letter.\footnotemark[264]\footnotetext[264]{\emph{See} Francis A. Gealogo, \emph{Looking for Clavería’s Children: Church, State, Power, and 
the Individual in Philippine Naming Systems During the Late Nineteenth Century}, \emph{in} Personal Names in Asia: History, Culture and 
Identity 63 (Zheng Yangwen \& Charles J.-H. Macdonald eds., NUS Press 2010).} A person’s name became a geographic tag readable by any 
clerk.\footnotemark[265]\footnotetext[265]{Id.} As businesses grew, displacing households as the locus of important economic activity, the 
state bestowed forms on them: the LLC, the partnership, and the corporation.\footnotemark[266]\footnotetext[266]{James Willard Hurst, The 
Legitimacy of the Business Corporation in the Law of the United States, 1780–1970 (1970).} Forms the state could see, name, and govern. 

AI agents present the state with a legibility crisis of the first order. But A-corps make AI agents legible. They do it, in the first 
instance, by solving the thin identity problem, connecting agents to something the state can already see and govern–namely, humans. 

But A-corps also solve the thick identity problem, making AI agents themselves legible to, and governable by, the state. They accomplish 
this without requiring anyone to know what AI agency “really” is, or to peer inside the black box of modern AI systems. Instead, A-corps 
create a legal-economic environment that forces AI entities to self-organize into legible forms. The state need not determine which AI 
entities share goals; it need only create stakes. Property creates stakes. Stakes create incentives. Incentives produce self-organization. 
And selection culls the AIs that fail to self-organize. In a slogan: A-corps are markets for personal identity. 

Doubtless, our proposal for granting legal rights and duties to non-human entities will sound radical to some readers. But it is no more 
radical than law’s prior grants of such entitlements to such entities. When legal systems first granted juridical personhood to corporations
 with “no soul to be damned, and no body to be kicked” many jurists were similarly 
scandalized.\footnotemark[267]\footnotetext[267]{\emph{See} John C. Coffee, Jr., \emph{“No Soul to Damn: No Body to Kick”: An Unscandalized 
Inquiry into the Problem of Corporate Punishment}, 79  Mich. L. Rev.  386 (1981).} Yet the corporate form has survived for hundreds of years
 because it is useful and, critically, governable. It creates a stable point for law to attach to businesses composed of ever-shifting 
swarms of humans, contracts, and capital. A-corps will extend the same logic further. 

The window for building the legal infrastructure needed to govern the AI economy is open now, while AI agents remain limited, and the swarms
 remain small. But it will not remain open indefinitely. The choices made now will shape the trajectory of a technology whose ultimate 
capabilities remain unknown. We should start building.

\end{document}